\documentclass[sigconf]{acmart}
\AtBeginDocument{%
  \providecommand\BibTeX{{%
    \normalfont B\kern-0.5em{\scshape i\kern-0.25em b}\kern-0.8em\TeX}}}

\copyrightyear{2023}
\acmYear{2023}
\setcopyright{rightsretained}
\acmConference[UIST '23]{The 36th Annual ACM Symposium on User Interface Software and Technology}{October 29-November 1, 2023}{San Francisco, CA, USA}
\acmBooktitle{The 36th Annual ACM Symposium on User Interface Software and Technology (UIST '23), October 29-November 1, 2023, San Francisco, CA, USA}
\acmDOI{10.1145/3586183.3606763}
\acmISBN{979-8-4007-0132-0/23/10}

\usepackage{multirow}
\usepackage{fancyvrb}
\usepackage{amsmath}
\usepackage{fontawesome5}
\usepackage{apple_emoji}
\usepackage[htt]{hyphenat}

\begin{document}

\title[Generative Agents]{Generative Agents: Interactive Simulacra of Human Behavior}

\author{Joon Sung Park}
\affiliation{%
  \institution{Stanford University}
  \city{Stanford}
  \country{USA}}
\email{joonspk@stanford.edu}

\author{Joseph C. O'Brien}
\affiliation{%
  \institution{Stanford University}
  \city{Stanford}
  \country{USA}}
\email{jobrien3@stanford.edu}

\author{Carrie J. Cai}
\affiliation{%
  \institution{Google Research}
  \city{Mountain View, CA}
  \country{USA}}
\email{cjcai@google.com}

\author{Meredith Ringel Morris}
\affiliation{%
  \institution{Google DeepMind}
  \city{Seattle, WA}
  \country{USA}}
\email{merrie@google.com}

\author{Percy Liang}
\affiliation{%
  \institution{Stanford University}
  \city{Stanford}
  \country{USA}}
\email{pliang@cs.stanford.edu}

\author{Michael S. Bernstein}
\affiliation{%
  \institution{Stanford University}
  \city{Stanford}
  \country{USA}}
\email{msb@cs.stanford.edu}

\renewcommand{\shortauthors}{J.S. Park, J.C. O'Brien, C.J. Cai, M.R. Morris, P. Liang, M.S. Bernstein}

\newcommand{\gentxt}[1]{{\color{darkgray}\small\fontfamily{}\selectfont
#1}}

\newcommand{\msb}[1]{{\color{blue}{MSB: #1}}}
\newcommand{\lp}[1]{{\color{magenta}{Lindsay: #1}}}
\newcommand{\mlg}[1]{{\color{red}{Mitchell: #1}}}
\newcommand{\joon}[1]{{\color{purple}{#1}}}
\newcommand{\still}[1]{{\color{orange}{#1}}}
\newcommand{\mirev}[1]{{\color{black}{#1}}}

\newcommand{\crdraft}[1]{{\color{black}{#1}}}

\begin{abstract}
  Believable proxies of human behavior can empower interactive applications ranging from immersive environments to rehearsal spaces for interpersonal communication to prototyping tools. In this paper, we introduce generative agents: computational software agents that simulate believable human behavior. Generative agents wake up, cook breakfast, and head to work; artists paint, while authors write; they form opinions, notice each other, and initiate conversations; they remember and reflect on days past as they plan the next day. To enable generative agents, we describe an architecture that extends a large language model to store a complete record of the agent's experiences using natural language, synthesize those memories over time into higher-level reflections, and retrieve them dynamically to plan behavior. We instantiate generative agents to populate an interactive sandbox environment inspired by The Sims, where end users can interact with a small town of twenty-five agents using natural language. In an evaluation, these generative agents produce believable individual and emergent social behaviors. For example, starting with only a single user-specified notion that one agent wants to throw a Valentine's Day party, the agents autonomously spread invitations to the party over the next two days, make new acquaintances, ask each other out on dates to the party, and coordinate to show up for the party together at the right time. We demonstrate through ablation that the components of our agent architecture—observation, planning, and reflection—each contribute critically to the believability of agent behavior. By fusing large language models with computational interactive agents, this work introduces architectural and interaction patterns for enabling believable simulations of human behavior.
\end{abstract}



\begin{teaserfigure}
  \includegraphics[width=\textwidth]{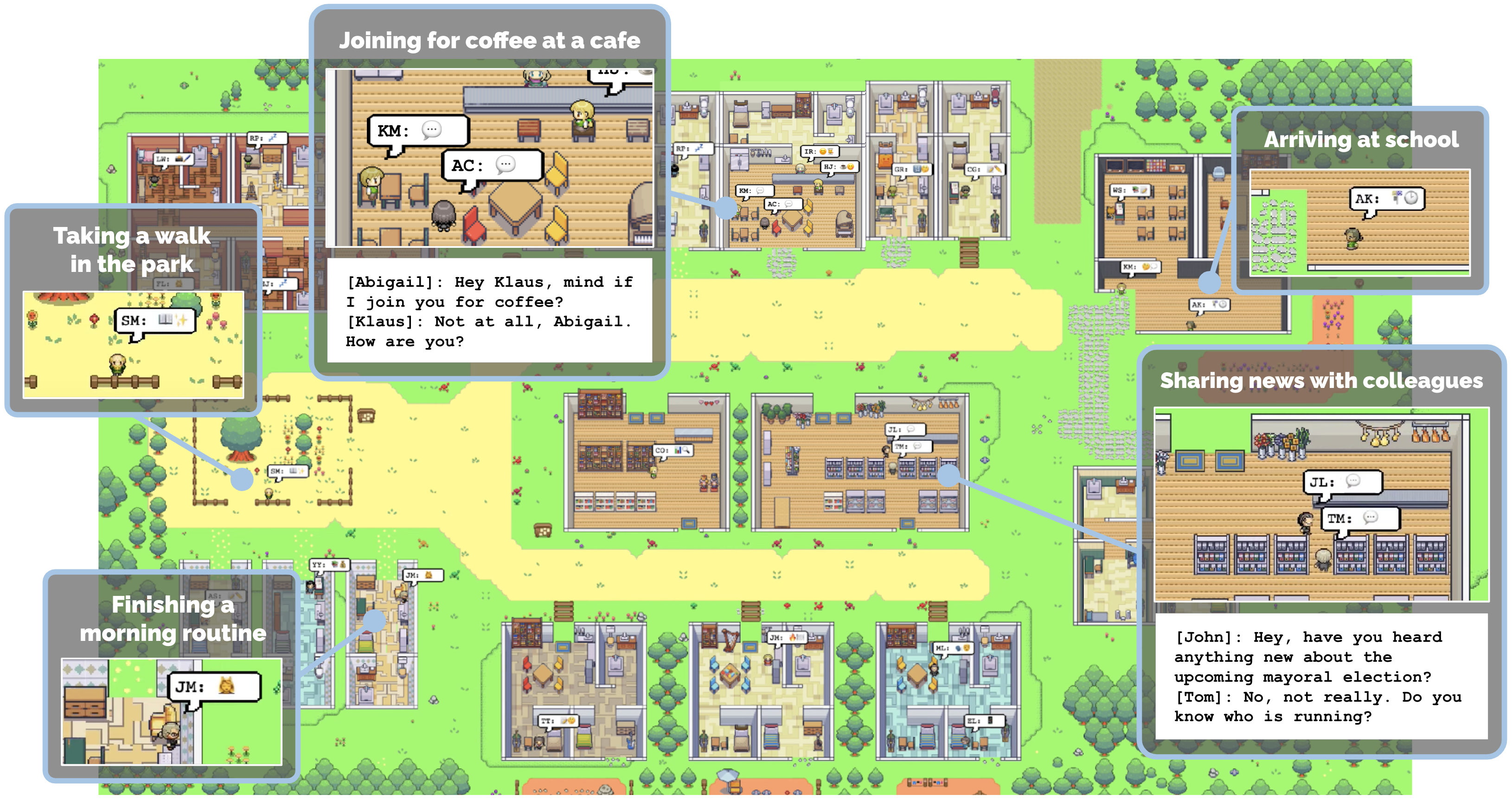}
  \caption{Generative agents are believable simulacra of human behavior for interactive applications. In this work, we demonstrate generative agents by populating a sandbox environment, reminiscent of The Sims, with twenty-five agents. Users can observe and intervene as agents plan their days, share news, form relationships, and coordinate group activities.}
  \Description{A screenshot of the game world populated by generative agents.}
  \label{fig:teaser}
\end{teaserfigure}

\maketitle

\section{Introduction}
How might we craft an interactive artificial society that reflects believable human behavior? From sandbox games such as The Sims to applications such as cognitive models~\cite{card1983psychology} and virtual environments~\cite{4_laird2001human, 22_bates1994role}, for over four decades, researchers and practitioners have envisioned computational agents that can serve as believable proxies of human behavior. In these visions, computationally-powered agents act consistently with their past experiences and react believably to their environments. Such simulations of human behavior could populate virtual spaces and communities with realistic social phenomena~\cite{29_dill2011game, 9_park2022socialsimulacra}, train people on how to handle rare yet difficult interpersonal situations~\cite{10_tambe1995intelligent, 11_jones1999automated, 20_hollan1984steamer}, test social science theories~\cite{binz2023using, 63_horton2023large}, craft model human processors for theory and usability testing~\cite{card1983psychology, john1996goms, hamalainen2023evaluating}, power ubiquitous computing applications~\cite{fast2016augur} and social robots~\cite{22_bates1994role, 41_Bledsoe1986Dream}, and underpin non-playable game characters~\cite{4_laird2001human, 6_riedl2012interactive} that can navigate complex human relationships in an open world. 

However, the space of human behavior is vast and complex~\cite{6_riedl2012interactive, 25_yannakakis2012game}. Despite striking progress in large language models~\cite{59_brown2020language} that can simulate human behavior at a single time point~\cite{9_park2022socialsimulacra,hamalainen2023evaluating}, fully general agents that ensure long-term coherence would be better suited by architectures that manage constantly-growing memories as new interactions, conflicts, and events arise and fade over time while handling cascading social dynamics that unfold between multiple agents. Success requires an approach that can \textit{retrieve} relevant events and interactions over a long period, \textit{reflect} on those memories to generalize and draw higher-level inferences, and apply that reasoning to create \textit{plans and reactions} that make sense in the moment and in the longer-term arc of the agent's behavior. 

In this paper, we introduce \textit{generative agents}---agents that draw on generative models to simulate believable human behavior---and demonstrate that they produce believable simulacra of both individual and emergent group behavior. Generative agents draw a wide variety of inferences about themselves, other agents, and their environment; they create daily plans that reflect their characteristics and experiences, act out those plans, react, and re-plan when appropriate; they respond when the end user changes their environment or commands them in natural language. For instance, generative agents turn off the stove when they see that their breakfast is burning, wait outside the bathroom if it is occupied, and stop to chat when they meet another agent they want to talk to.\footnote{\crdraft{When referring to generative agents engaging in actions or going to places, this is a shorthand for readability and not a suggestion that they are engaging in human-like agency. The behaviors of our agents, akin to animated Disney characters, aim to create a sense of believability, but they do not imply genuine agency.}} A society full of generative agents is marked by emergent social dynamics where new relationships are formed, information diffuses, and coordination arises across agents. 

To enable generative agents, we describe an agent architecture that stores, synthesizes, and applies relevant memories to generate believable behavior using a large language model. Our architecture comprises three main components. The first is the \textit{memory stream}, a long-term memory module that records, in natural language, a comprehensive list of the agent's experiences. A memory retrieval model combines relevance, recency, and importance to surface the records needed to inform the agent's moment-to-moment behavior. The second is \textit{reflection}, which synthesizes memories into higher-level inferences over time, enabling the agent to draw conclusions about itself and others to better guide its behavior. The third is \textit{planning}, which translates those conclusions and the current environment into high-level action plans and then recursively into detailed behaviors for action and reaction. These reflections and plans are fed back into the memory stream to influence the agent's future behavior.

This architecture suggests applications in multiple domains, from role-play and social prototyping to virtual worlds and games. In social role-play scenarios (e.g., interview preparation), a user could safely rehearse difficult, conflict-laden conversations. When prototyping social platforms, a designer could go beyond temporary personas to prototype dynamic, complex interactions that unfold over time. For this paper, we focus on the ability to create a small, interactive society of agents inspired by games such as The Sims.\footnote{A demonstration of an actual simulation of the generative agent society can be viewed at the following link: \url{https://reverie.herokuapp.com/UIST_Demo/}. A public repository for the simulation code is located here: \url{https://github.com/joonspk-research/generative_agents}} By connecting our architecture to the ChatGPT large language model~\cite{openai_chatgpt}, we manifest a society of twenty-five agents in a game environment. End users can observe and interact with these agents. If an end user or developer wanted the town to host an in-game Valentine's Day party, for example, traditional game environments would require scripting tens of characters' behavior manually. We demonstrate that, with generative agents, it is sufficient to simply tell one agent that she wants to throw a party. Despite many potential points of failure---the party planner must remember to invite other agents to the party, attendees must remember the invitation, those who remember must decide to actually show up, and more---our agents succeed. They spread the word about the party and then show up, with one agent even asking another on a date to the party, all from a single user-generated seed suggestion.

We conducted two evaluations of generative agents: a controlled evaluation to test whether the agents produce believable individual behaviors in isolation, and an end-to-end evaluation where the agents interacted with each other in open-ended ways over two days of game time to understand their stability and emergent social behaviors. In the technical evaluation, we leverage a methodological opportunity to evaluate an agent's knowledge and behavior by ``interviewing'' it in natural language to probe the agents' ability to stay in character, remember, plan, react, and reflect accurately. We compared several ablations that limit agents' access to memory, reflection, and planning. We observe that each of these components is critical to strong performance across these interview tasks. Across the technical and end-to-end evaluation, the most common errors arose when the agent failed to retrieve relevant memories, fabricated embellishments to the agent's memory, or inherited overly formal speech or behavior from the language model.

In sum, this paper makes the following contributions:
\begin{itemize}
    \item \textit{Generative agents}, believable simulacra of human behavior that are dynamically conditioned on agents' changing experiences and environment.
    \item A novel architecture that makes it possible for generative agents to remember, retrieve, reflect, interact with other agents, and plan through dynamically evolving circumstances. The architecture leverages the powerful prompting capabilities of large language models and supplements those capabilities to support longer-term agent coherence, the ability to manage dynamically evolving memory, and recursively produce higher-level reflections.
    \item Two evaluations, a controlled evaluation and an end-to-end evaluation, that establish causal effects of the importance of components of the architecture, as well as identify breakdowns arising from, e.g., improper memory retrieval. 
    \item Discussion of the opportunities and ethical and societal risks of generative agents in interactive systems. We argue that these agents should be tuned to mitigate the risk of users forming parasocial relationships, logged to mitigate risks stemming from deepfakes and tailored persuasion, and applied in ways that complement rather than replace human stakeholders in design processes.
\end{itemize}

\section{Related Work}
In this section, we reflect on the prior literature in human-AI interaction and situate, within its canon, the agenda of building believable proxies of human behavior. This agenda, once hailed as a north star in the interaction, game, and artificial intelligence communities~\cite{4_laird2001human, 6_riedl2012interactive, 21_riedl2005objective, 22_bates1994role}, has remained challenging due to the complexity of human behavior~\cite{23_brooks2000cog, 25_yannakakis2012game}. We synthesize this research to suggest that large language models, though not sufficient by themselves, open up a new angle for creating believable agents when leveraged using the appropriate architecture.

\subsection{Human-AI Interaction}
Interactive artificial intelligence systems aim to combine human insights and capabilities in computational artifacts that can augment their users ~\cite{amershi2014power, fails2003interactive}. A long line of work has explored ways to enable users to interactively specify model behavior. For instance, Crayons demonstrated an early vision of interactive machine learning, allowing non-expert users to train classifiers~\cite{fails2003interactive}. Further work helped to articulate how end users might describe their classification goals to the system through examples~\cite{Fogarty2008} or demonstration~\cite{fiebrink2010wekinator}. Recent advancements have extended these explorations to deep learning~\cite{lam2023model} and prompt-based authoring~\cite{jiang_PromptMaker,62_Wu2021AIChains,liu2022opal}.

Meanwhile, a persistent thread of research has advanced the case for language- and agent-based interaction in human-computer interaction. Formative work such as SHRDLU~\cite{winograd1971procedures} and ELIZA~\cite{weizenbaum1966eliza} demonstrated the opportunities and the risks associated with natural language interaction with computing systems. As research progressed, it became evident that autonomous agents could offer new metaphors for delegation and interaction~\cite{MaesPattie1995}, but the boundaries of delegation between humans and agents have remained the subject of ongoing debate and refinement~\cite{shneiderman1997direct,horvitz1999principles,shneiderman2022human}. Recently, this technology has reached a level of stability that enables agents to interact via natural language in large and complex online social environments (e.g.,~\cite{Krishna2022}). Natural language interaction provides a novel modality that can enhance user abilities in domains such as photo editing~\cite{laput2013pixeltone, fourney2011query, adar2014command} and code editing~\cite{rong2016codemend}.

We convene these threads of work to show that we can now create agents that proxy human behavior for interactive systems, and interact with them using natural language. In doing so, this work reopens the door to examining foundational human-computer interaction questions around cognitive models such as GOMS and Keystroke-Level Model (KLM)~\cite{card1983psychology,card1980keystroke}, around prototyping tools~\cite{9_park2022socialsimulacra}, and around ubiquitous computing applications~\cite{weiser1991computer,dey2001understanding,fast2016augur}.

\subsection{Believable Proxies of Human Behavior}
Prior literature has described \textit{believability}, or \textit{believable agents}, as a central design and engineering goal. Believable agents are designed to provide an illusion of life and present a facade of realism in the way they appear to make decisions and act on their own volition, similar to the characters in Disney movies~\cite{22_bates1994role, 42_Thomas1981Disney}. These agents can populate and perceive an open world environment like the one we inhabit~\cite{4_laird2001human, 22_bates1994role}, and strive to behave in ways that exhibit emergent behaviors grounded in social interactions with users or other agents with the aim of becoming believable proxies of our behavior in hypothetical simulations of individuals and communities~\cite{26_mccoy2012prom, 43_Burkinshaw2009Alice, 44_Francis2010Minecraft}. Historically, these agents were developed in the context of intelligent game non-player characters (NPCs)~\cite{4_laird2001human, 6_riedl2012interactive}. Creating NPCs with believable behavior, if possible, could enhance player experiences in games and interactive fictions by enabling emergent narratives~\cite{1_swartout2006toward, 2_aylett1999narrative, 3_brenner2010creating, ibister2000consistency} and social interactions with the agents~\cite{5_zubek2002towards}. However, more importantly, game worlds provide increasingly realistic representations of real-world affordances, and as observed by Laird and van Lent in 2001, these simulated worlds offer accessible testbeds for developers of believable agents to finesse the agents' cognitive capabilities without worrying about implementing robotics in the real world or creating simulation environments from scratch~\cite{4_laird2001human, 6_riedl2012interactive}.
 
A diverse set of approaches to creating believable agents emerged over the past four decades.
In implementation, however, these approaches often simplified the environment or dimensions of agent behavior to make the effort more manageable~\cite{23_brooks2000cog, 39_Minsky1970Draft}. Rule-based approaches, such as finite-state machines~\cite{30_NEURIPS2021_86e8f7ab, 32_Umarov2012Believable} and behavior trees~\cite{45_Knafla2011Introduction, 46_Pillosu2009Coordinating, 47_Hecker2011Spore} account for the brute force approach of human-authoring the agent's behavior~\cite{26_mccoy2012prom}. They provide a straightforward way of creating simple agents that is still the most dominant approach today~\cite{37_McCoy2009Comme, 24_miyashita2017developing, 25_yannakakis2012game}, and can even handle rudimentary social interactions, as shown in games such as Mass Effect~\cite{48_BioWare2007Mass} and The Sims~\cite{49_ElectronicArts2009Sims} series. Nonetheless, manually crafting behavior that can comprehensively address the breadth of possible interactions in an open world is untenable. This means that the resulting agent behaviors may not fully represent the consequences of their interactions~\cite{26_mccoy2012prom, 27_mccoy2011prom, 28_mccoy2011comme}, and cannot perform new procedures that were not hard-coded in their script~\cite{30_NEURIPS2021_86e8f7ab, 32_Umarov2012Believable}. On the other hand, prevalent learning-based approaches for creating believable agents, such as reinforcement learning, have overcome the challenge of manual authoring by letting the agents learn their behavior, and have achieved superhuman performance in recent years in games such as AlphaStar for Starcraft~\cite{50_Vinyals2019Grandmaster} and OpenAI Five for Dota 2~\cite{51_Berner2019Dota}. However, their success has largely taken place in adversarial games with readily definable rewards that a learning algorithm can optimize for. They have not yet addressed the challenge of creating believable agents in an open world~\cite{30_NEURIPS2021_86e8f7ab, 24_miyashita2017developing, 38_Hausknecht2020Interactive}.

Cognitive architectures in computation, pioneered by Newell, aimed to build the infrastructure for supporting a comprehensive set of cognitive functions~\cite{14_newell1990unified} that suited the all-encompassing nature of believable agents held in its original vision. They fueled some of the earliest examples of believable agents. For instance, Quakebot-SOAR~\cite{54_Laird2000It} and ICARUS~\cite{33_LangleyIcarus, 34_Choi2021Believable} generated NPCs in first-person shooter games, while TacAir-SOAR~\cite{55_Pew1998Modeling} generated pilots in aerial combat training simulations. The architectures used by these agents differed (Quakebot- and TacAir-SOAR relied on SOAR~\cite{56_Laird2012Soar}, while ICARUS relied on its own variation that was inspired by SOAR and ACT-R~\cite{57_Anderson1993Rules}), but they shared the same underlying principle~\cite{58_Laird2017Standard}. They maintained short-term and long-term memories, filled these memories with symbolic structures, and operated in perceive-plan-act cycles, dynamically perceiving the environment and matching it with one of the manually crafted action procedures~\cite{32_Umarov2012Believable, 36_Laird2001Knows}. Agents created using cognitive architectures aimed to be generalizable to most, if not all, open world contexts and exhibited robust behavior for their time. However, their space of action was limited to manually crafted procedural knowledge, and they did not offer a mechanism through which the agents could be inspired to seek new behavior. As such, these agents were deployed mostly in non-open world contexts such as first-person shooter games~\cite{54_Laird2000It, 34_Choi2021Believable} or blocks worlds~\cite{33_LangleyIcarus}.

Today, creating believable agents as described in its original definition remains an open problem~\cite{25_yannakakis2012game, 6_riedl2012interactive}. Many have moved on, arguing that although current approaches for creating believable agents might be cumbersome and limited, they are good enough to support existing gameplay and interactions~\cite{25_yannakakis2012game, 52_Champandard2012Tutorial, 53_Nareyek2007Game}. Our argument is that large language models offer an opportunity to re-examine these questions, provided that we can craft an effective architecture to synthesize memories into believable behavior. We offer a step toward such an architecture in this paper.

\subsection{Large Language Models and Human Behavior}
Generative agents leverage a large language model to power their behavior. The key observation is that large language models encode a wide range of human behavior from their training data~\cite{59_brown2020language, 60_bommasani2022opportunities}. If prompted with a narrowly defined context, the models can be used to generate believable behavior. Recent work has demonstrated the efficacy of this approach. For instance, social simulacra used a large language model to generate users that would populate new social computing systems to prototype their emergent social dynamics~\cite{9_park2022socialsimulacra}. This approach used a prompt chain~\cite{61_Wu2022PromptChainer, 62_Wu2021AIChains} to generate short natural language descriptions of personas and their behaviors as they appear in the system being prototyped. Other empirical studies have replicated existing social science studies~\cite{63_horton2023large}, political surveys~\cite{64_Sorensen_2022}, and generated synthetic data~\cite{hamalainen2023evaluating}. Large language models have also been used to generate interactive human behavior for users to engage with. In gaming, for instance, these models have been employed to create interactive fiction~\cite{65_Freiknecht2020Procedural} and text adventure games~\cite{66_callison-burch-etal-2022-dungeons}. With their ability to generate and decompose action sequences, large language models have also been used in planning robotics tasks~\cite{67_huang2022inner}. For example, when presented with a task, such as picking up a bottle, the model is prompted to break down the task into smaller action sequences, such as heading to the table where the bottle is located and picking it up.

We posit that, based on the work summarized above, large language models can become a key ingredient for creating believable agents. The existing literature largely relies on what could be considered first-order templates that employ few-shot prompts~\cite{68_Gao2020Making, 69_Liu2021What} or chain-of-thought prompts~\cite{70_wei2023chainofthought}. These templates are effective in generating behavior that is conditioned solely on the agent's current environment (e.g., how would a troll respond to a given post, what actions would a robot need to take to enter a room given that there is a door). However, believable agents require conditioning not only on their current environment but also on a vast amount of past experience, which is a poor fit (and as of today, impossible due to the underlying models' limited context window) using first-order prompting. Recent studies have attempted to go beyond first-order prompting by augmenting language models with a static knowledge base and an information retrieval scheme~\cite{khattab2023demonstratesearchpredict} or with a simple summarization scheme~\cite{wu2021recursively}. This paper extends these ideas to craft an agent architecture that handles retrieval where past experience is dynamically updated at each time step and mixed with agents' current context and plans, which may either reinforce or contradict each other.
\begin{figure*}[tb]
  \centering
  \includegraphics[width=0.98\textwidth]{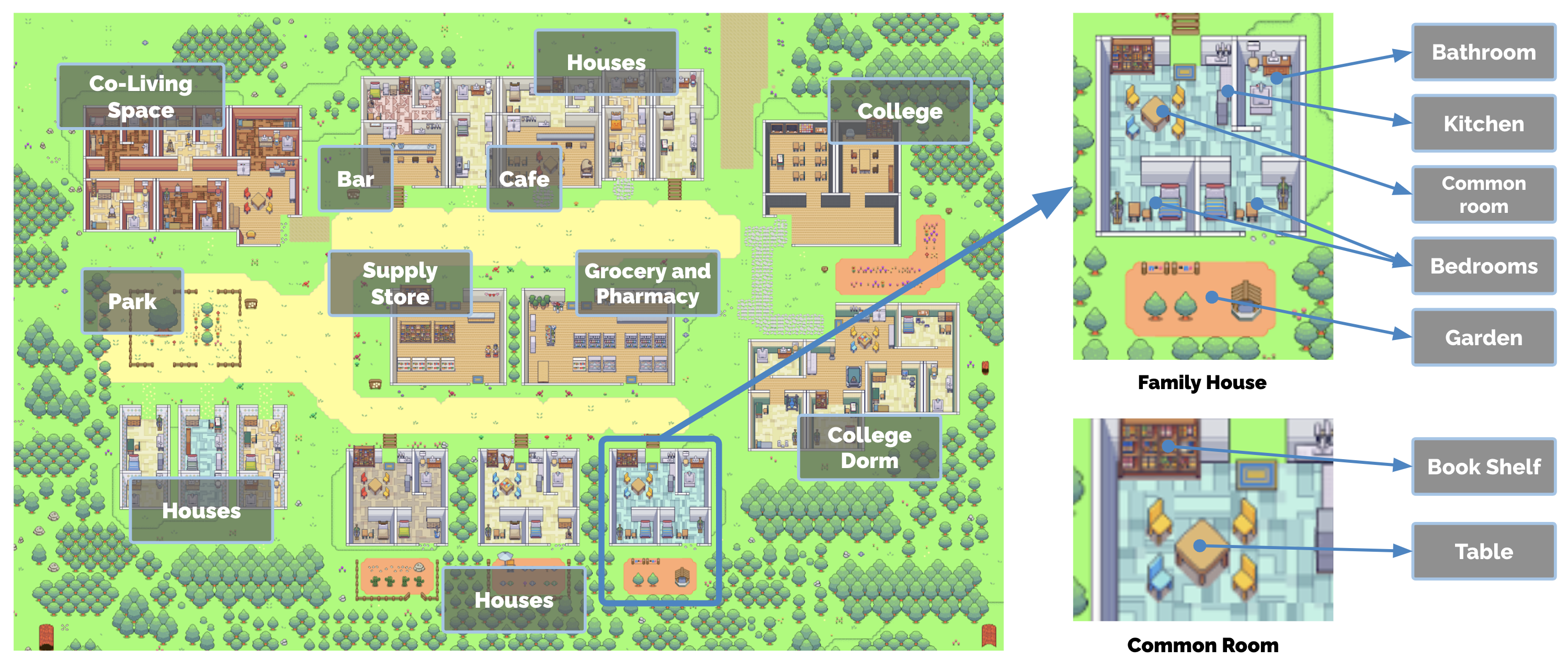}
  \caption{The Smallville sandbox world, with areas labeled. The root node describes the entire world, children describe areas (e.g., houses, cafe, stores), and leaf nodes describe objects (e.g., table, bookshelf). Agents remember a subgraph that reflects the parts of the world they have seen, maintaining the state of those parts as they observed them.}
  \Description{The labeled map of the game world.}
  \label{fig:game_world}
\end{figure*}

\section{Generative Agent Behavior and Interaction}\label{sec:behavior}
To illustrate the affordances of generative agents, we instantiate them as characters in a simple sandbox world reminiscent of The Sims~\cite{49_ElectronicArts2009Sims}. This sprite-based sandbox game world, Smallville, evokes a small town environment. In this section, we will walk through the affordances and interactions with generative agents in Smallville and describe how the agents behave within it. Then, in Section~\ref{sec:architecture}, we will introduce our generative agent architecture that powers these affordances and interactions. In Section~\ref{sec:implementation}, we will describe the implementation of the sandbox environment and how the agents interact with the underlying engine of the sandbox world.

\subsection{Agent Avatar and Communication}\label{sec:avatar}
A community of 25 unique agents inhabits Smallville. Each agent is represented by a simple sprite avatar. We authored one paragraph of natural language description to depict each agent's identity, including their occupation and relationship with other agents, as seed memories. For example, John Lin has the following description:
\begin{quote}
{\small
\texttt{John Lin is a pharmacy shopkeeper at the Willow Market and Pharmacy who loves to help people. He is always looking for ways to make the process of getting medi\-cation easier for his customers; John Lin is living with his wife, Mei Lin, who is a college professor, and son, Eddy Lin, who is a student studying music theory; John Lin loves his family very much; John Lin has known the old couple next-door, Sam Moore and Jennifer Moore, for a few years; John Lin thinks Sam Moore is a kind and nice man; John Lin knows his neighbor, Yuriko Yamamoto, well; John Lin knows of his neighbors, Tamara Taylor and Carmen Ortiz, but has not met them before; John Lin and Tom Moreno are colleagues at The Willows Market and Pharmacy; John Lin and Tom Moreno are friends and like to discuss local politics together; John Lin knows the Moreno family somewhat well — the husband Tom Moreno and the wife Jane Moreno.}
}
\end{quote}
Each semicolon-delimited phrase is entered into the agent's initial memory as memories at the start of the simulation. 

\subsubsection{Inter-Agent Communication}
The agents interact with the world by their actions, and with each other through natural language. At each time step of the sandbox engine, the agents output a natural language statement describing their current action, such as ``Isabella Rodriguez is writing in her journal'', ``Isabella Rodriguez is checking her emails'', ``Isabella Rodriguez is talking with her family on the phone'', or ``Isabella Rodriguez is getting ready for bed.'' This statement is then translated into concrete movements that affect the sandbox world. The action is displayed on the sandbox interface as a set of emojis, providing an abstract representation of the action from an overhead view. To achieve this, the system utilizes a language model to translate the action into a set of emojis, which appear above each avatar's head in a speech bubble. For example, ``Isabella Rodriguez is writing in her journal'' is displayed as \raisebox{-0.1cm}{\protect\includegraphics[width=0.3cm]{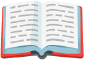}}\raisebox{-0.1cm}{\protect\includegraphics[width=0.3cm]{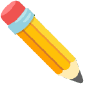}}, while ``Isabella Rodriguez is checking her emails'' appears as \raisebox{-0.1cm}{\protect\includegraphics[width=0.3cm]{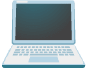}}\,\raisebox{-0.1cm}{\protect\includegraphics[width=0.3cm]{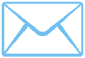}}. The complete natural language description of the action can be accessed by clicking on the agent's avatar.

Agents communicate with each other in full natural language. They are aware of other agents in their local area, and the generative agent architecture determines whether they walk by or engage in conversation. Here, a sample in the middle of a conversation between the agents Isabella Rodriguez and Tom Moreno about the upcoming election:\footnote{We note that the conversational style of these agents can feel overly formal, likely a result of instruction tuning in the underlying models. We expect that the writing style will be better controllable in future language models.}
\begin{quote}
    \gentxt{
        \textbf{Isabella}: I’m still weighing my options, but I’ve been discussing the election with Sam Moore. What are your thoughts on him? \\
        \textbf{Tom}: To be honest, I don’t like Sam Moore. I think he’s out of touch with the community and doesn’t have our best interests at heart.
    }
\end{quote}

\subsubsection{User Controls}
The user communicates with the agent through natural language by specifying a persona that the agent should perceive them as. For example, if the user specifies that they are a news ``reporter'' and asks about the upcoming election by saying, ``Who is running for office?'', the John agent replies:
\begin{quote}
    \gentxt{
        \textbf{John}: My friends Yuriko, Tom and I have been talking about the upcoming election and discussing the candidate Sam Moore. We have all agreed to vote for him because we like his platform.
    }
\end{quote}
To directly command one of the agents, the user takes on the persona of the agent’s ``inner voice''---this makes the agent more likely to treat the statement as a directive. For instance, when told “You are going to run against Sam in the upcoming election” by a user as John's inner voice, John decides to run in the election and shares his candidacy with his wife and son.

\subsection{Environmental Interaction}
Smallville features the common affordances of a small village, including a cafe, bar, park, school, dorm, houses, and stores. It also defines subareas and objects that make those spaces functional, such as a kitchen in a house and a stove in the kitchen (Figure~\ref{fig:game_world}). All spaces serving as agents’ primary living quarters feature a bed, desk, closet, shelf, as well as a bathroom and a kitchen.\footnote{This environment design is not the focus of our work, so we generated this environment manually, not automatically. Future work can continue to expand the richness of the agents' environments.}

Agents move around Smallville as one would in a simple video game, entering and leaving buildings, navigating its map, and approaching other agents. Agent movements are directed by the generative agent architecture and the sandbox game engine: when the model dictates that the agent will move to a location, we calculate a walking path to the destination in the Smallville environment, and the agent begins moving. In addition, users can also enter the sandbox world of Smallville as an agent operating within it. The agent that the user embodies can be an agent already present in the world, such as Isabella and John, or it can be an outside visitor with no prior history in Smallville. The inhabitants of Smallville will treat the user-controlled agent no differently than they treat each other. They recognize its presence, initiate interactions, and remember its behavior before forming opinions about it.

Users and agents can influence the state of the objects in this world, much like in sandbox games such as The Sims. For example, a bed can be occupied when an agent is sleeping, and a refrigerator can be empty when an agent uses up the ingredients to make breakfast. End users can also reshape an agent's environment in Smallville by rewriting the status of objects surrounding the agent in natural language. For instance, when Isabella is making breakfast in the morning, the user can change the status of the kitchen stove from “turned on” to “burning” by inputting a command to the system that chooses the object and illustrates its new status, like this: ``<Isabella's apartment: kitchen: stove> is burning.'' Isabella will notice this in the next moment and go to turn off the stove and remake her breakfast. Likewise, if the user sets the status of Isabella's shower to “leaking water” when she enters the bathroom, she will gather tools from her living room and try to fix the leak. 

\begin{figure*}[tb]
  \centering
  \includegraphics[width=0.95\textwidth]{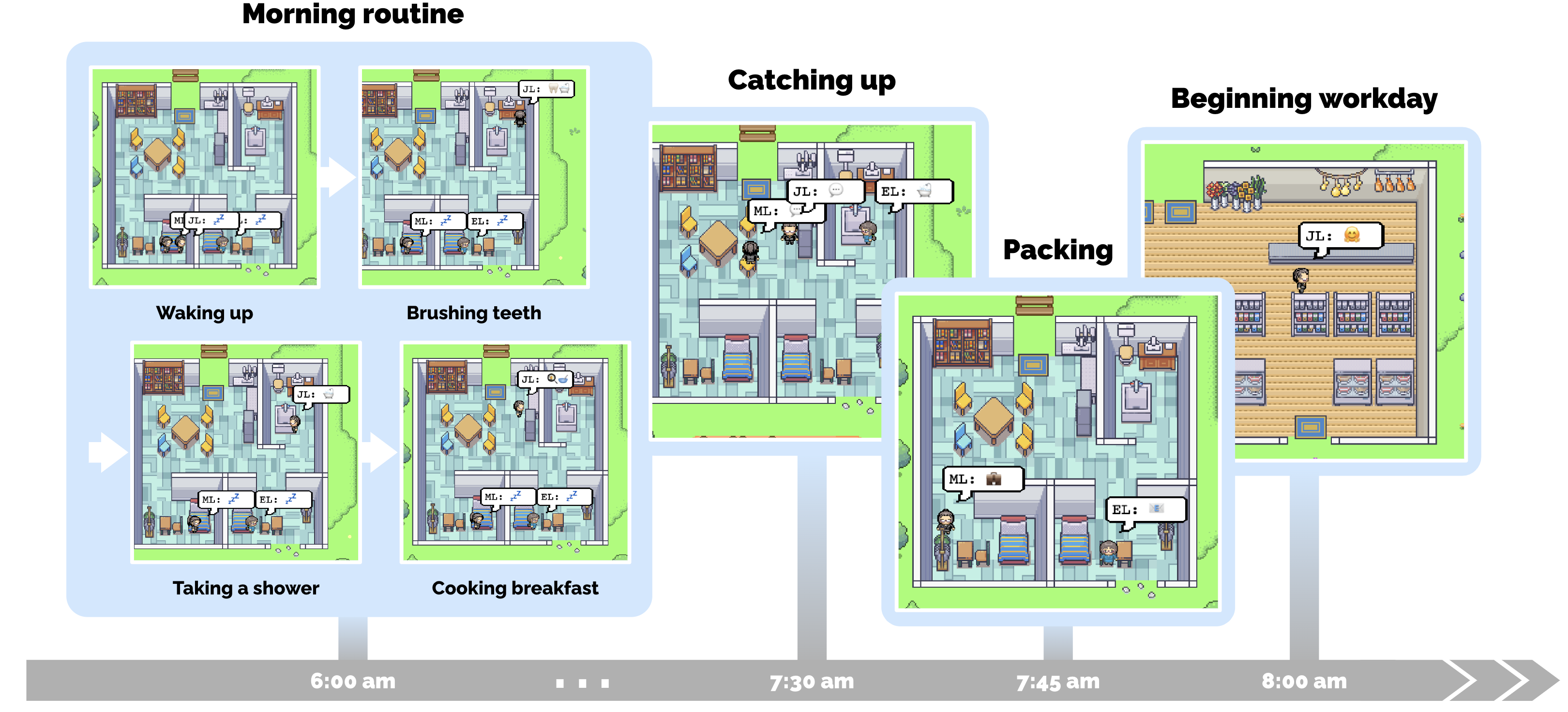}
  \caption{A morning in the life of a generative agent, John Lin. John wakes up around 6 am and completes his morning routine, which includes brushing his teeth, taking a shower, and eating breakfast. He briefly catches up with his wife, Mei, and son, Eddy, before heading out to begin his workday.}
  \Description{The Lin family's morning routines.}
  \label{fig:Lin_family_morning}
\end{figure*}

\begin{figure}[tb]
  \centering
  \includegraphics[width=0.75\columnwidth]{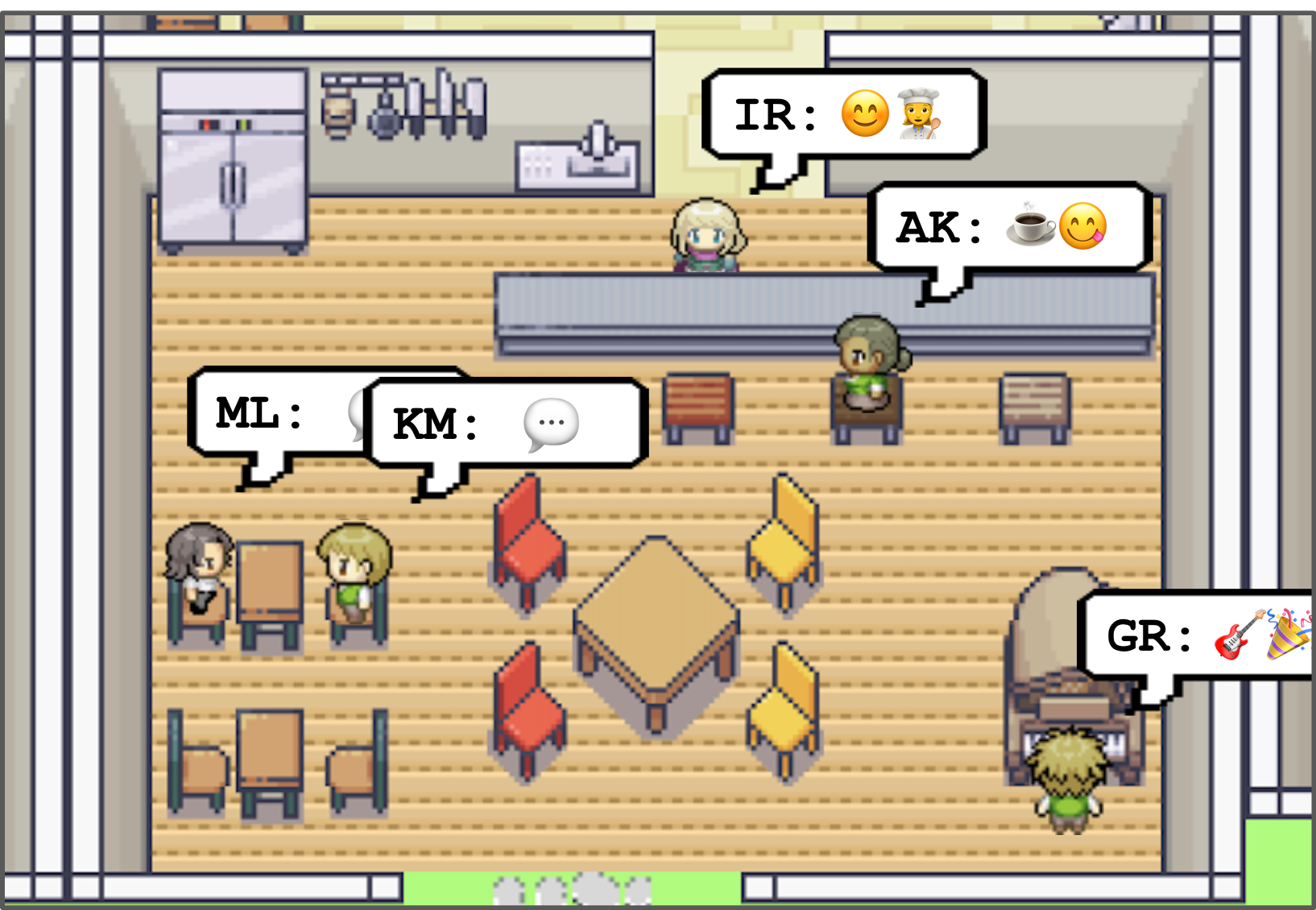}
  \caption{At the beginning of the simulation, one agent is initialized with an intent to organize a Valentine's Day party. Despite many possible points of failure in the ensuing chain of events---agents might not act on that intent, might forget to tell others, might not remember to show up---the Valentine's Day party does, in fact, occur, with a number of agents gathering and interacting.}
  \Description{The Valentine's day party at Hobbs Cafe.}
  \label{fig:valentine}
\end{figure}

\subsection{Example ``Day in the Life''}
Starting from the single-paragraph description, generative agents begin planning their days. As time passes in the sandbox world, their behaviors evolve as these agents interact with each other and the world, building memories and relationships, and coordinating joint activities.

We demonstrate the behavior of generative agents by tracing the output of our system over the course of one day for the agent John Lin (Figure~\ref{fig:Lin_family_morning}). In the Lin family, John is the first to wake up at 7 a.m. He brushes his teeth, takes a shower, gets dressed, eats breakfast, and checks the news at the dining table in the living room. At 8 a.m., Eddy follows suit, rushing out of bed to prepare for his classes. He catches up with John just as he is about to head out:
\begin{quote}
    \gentxt{
        \textbf{John}: Good morning Eddy. Did you sleep well? \\
        \textbf{Eddy}: Good morning dad. Yeah, I slept great. \\
        \textbf{John}: That's good. What are you working on today? \\
        \textbf{Eddy}: I'm working on a new music composition for my class. It's due this week, so I'm trying to get it finished. But I'm having so much fun with it! \\
        \textbf{John}: That sounds great!}  
\end{quote}
Soon after Eddy heads out, Mei wakes up and joins John. Mei asks about her son, and John recalls the conversation they just had:
\begin{quote}
    \gentxt{
        \textbf{Mei}: Did Eddy already leave for school? \\
        \textbf{John}: Yes, he just left. He's working on a music composition for his class. \\
        \textbf{Mei}: Oh, that's wonderful! Tell me about it. \\
        \textbf{John}: I think he's really enjoying it! He said he's having a lot of fun with it. \\
        \textbf{Mei}: That's great! I'm so proud of him.
    }
\end{quote}
As they finish their conversation, Mei and John pack up. Mei goes on to teach and work on her research paper, while John opens his pharmacy counter at the Willow Market and Pharmacy by 9 a.m. 

\begin{figure*}[tb]
  \centering
  \includegraphics[width=0.88\textwidth]{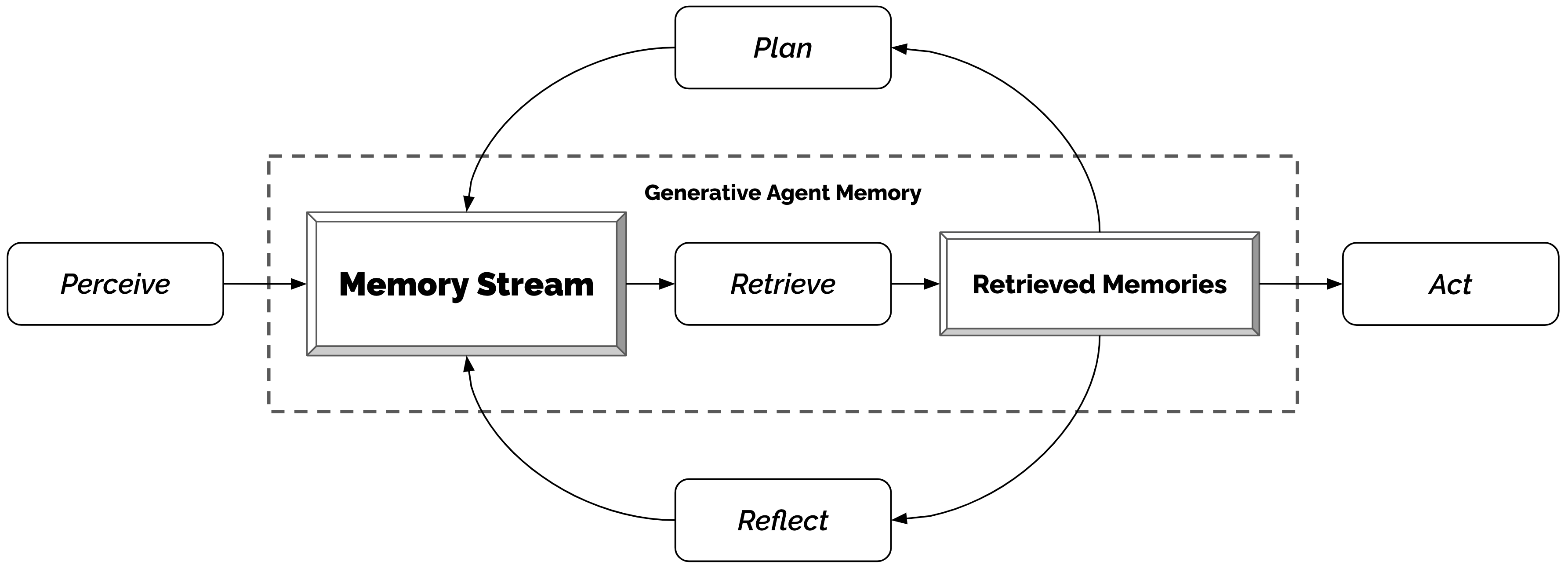}
  \caption{Our generative agent architecture. Agents perceive their environment, and all perceptions are saved in a comprehensive record of the agent's experiences called the memory stream. Based on their perceptions, the architecture retrieves relevant memories and uses those retrieved actions to determine an action. These retrieved memories are also used to form longer-term plans and create higher-level reflections, both of which are entered into the memory stream for future use.}
  \Description{The architecture of generative agents.}
  \label{fig:architecture}
\end{figure*}

\subsection{Emergent Social Behaviors}
By interacting with each other, generative agents in Smallville exchange information, form new relationships, and coordinate joint activities. Extending prior work~\cite{9_park2022socialsimulacra}, these social behaviors are emergent rather than pre-programmed. 

\subsubsection{Information Diffusion} 
As agents notice each other, they may engage in dialogue---as they do so, information can spread from agent to agent. For instance, in a conversation between Sam and Tom at the grocery store, Sam tells Tom about his candidacy in the local election:
\begin{quote}
    \gentxt{
        \textbf{Sam}: Hey Tom, how's it going? \\
        \textbf{Tom}: Good, thanks. What's up? \\
        \textbf{Sam}: Well, I wanted to talk to you about something. I'm actually running for mayor in the upcoming local election. \\
        \textbf{Tom}: Really? That's great news! Why are you running? \\
        \textbf{Sam}: I've been involved in local politics for years now, and I really feel like I can make a difference in our community. Plus, I want to bring new ideas and perspectives to the table. 
    }
\end{quote}
Later that day, after Sam left, Tom and John, who heard the news from another source, discuss Sam's chances of winning the election:
\begin{quote}
    \gentxt{
        \textbf{John}: I heard that Sam Moore is running for mayor in the local election. Do you think he has a good chance of winning? \\
        \textbf{Tom}: I do think he has a good chance. He's been working hard in the community and I think he will get a lot of support. What do you think? \\
        \textbf{John}: I think it's great that he's running. I'm curious to see how the election will turn out.
    }
\end{quote}
Gradually, Sam's candidacy becomes the talk of the town, with some supporting him and others remaining undecided. 

\subsubsection{Relationship Memory} 
Agents in Smallville form new relationships over time and remember their interactions with other agents. For example, at the start, Sam does not know Latoya Williams. While taking a walk in Johnson Park, Sam runs into Latoya, and they introduce themselves. Latoya mentions that she is working on a photography project: \gentxt{``I'm here to take some photos for a project I'm working on.''} In a later interaction, Sam's interactions with Latoya indicate a memory of that interaction, as he asks \gentxt{``Hi, Latoya. How is your project going?''} and she replies \gentxt{``Hi, Sam. It’s going well!''}

\subsubsection{Coordination} 
Generative agents coordinate with each other. Isabella Rodriguez, at Hobbs Cafe, is initialized with an intent to plan a Valentine's Day party from 5 to 7 p.m. on February 14th. From this seed, the agent proceeds to invite friends and customers when she sees them at Hobbs Cafe or elsewhere. Isabella then spends the afternoon of the 13th decorating the cafe for the occasion. Maria, a frequent customer and close friend of Isabella's, arrives at the cafe. Isabella asks for Maria's help in decorating for the party, and Maria agrees. Maria's character description mentions that she has a crush on Klaus. That night, Maria invites Klaus, her secret crush, to join her at the party, and he gladly accepts.

On Valentine's Day, five agents, including Klaus and Maria, show up at Hobbs Cafe at 5 pm, and they enjoy the festivities (Figure~\ref{fig:valentine}). In this scenario, the end user only set Isabella's initial intent to throw a party and Maria's crush on Klaus: the social behaviors of spreading the word, decorating, asking each other out, arriving at the party, and interacting with each other at the party were initiated by the agent architecture.

\section{Generative Agent Architecture}\label{sec:architecture}
Generative agents aim to provide a framework for behavior in an open world: one that can engage in interactions with other agents and react to changes in the environment. Generative agents take their current environment and past experiences as input and generate behavior as output. Underlying this behavior is a novel agent architecture that combines a large language model with mechanisms for synthesizing and retrieving relevant information to condition the language model's output. Without these mechanisms, large language models can output behavior, but the resulting agents may not react based on the agent's past experiences, may not make important inferences, and may not maintain long-term coherence. Challenges with long-term planning and coherence remain~\cite{bubeck2023sparks} even with today's most performant models such as GPT-4. Because generative agents produce large streams of events and memories that must be retained, a core challenge of our architecture is to ensure that the most relevant pieces of the agent's memory are retrieved and synthesized when needed.

At the center of our architecture is the memory stream, a database that maintains a comprehensive record of an agent's experience. From the memory stream, records are retrieved as relevant to plan the agent's actions and react appropriately to the environment. Records are recursively synthesized into higher- and higher-level reflections that guide behavior. Everything in the architecture is recorded and reasoned over as a natural language description, allowing the architecture to leverage a large language model.

Our current implementation utilizes the gpt3.5-turbo version of ChatGPT~\cite{openai_chatgpt}. We expect that the architectural basics of generative agents---memory, planning, and reflection---will likely remain the same as language models improve. Newer language models (e.g., GPT-4) will continue to expand the expressive power and performance of the prompts that underpin generative agents. As of writing, however, GPT-4's API was invitation-only, so our agents use ChatGPT.

\begin{figure*}[tb]
  \centering
  \includegraphics[width=0.86\textwidth]{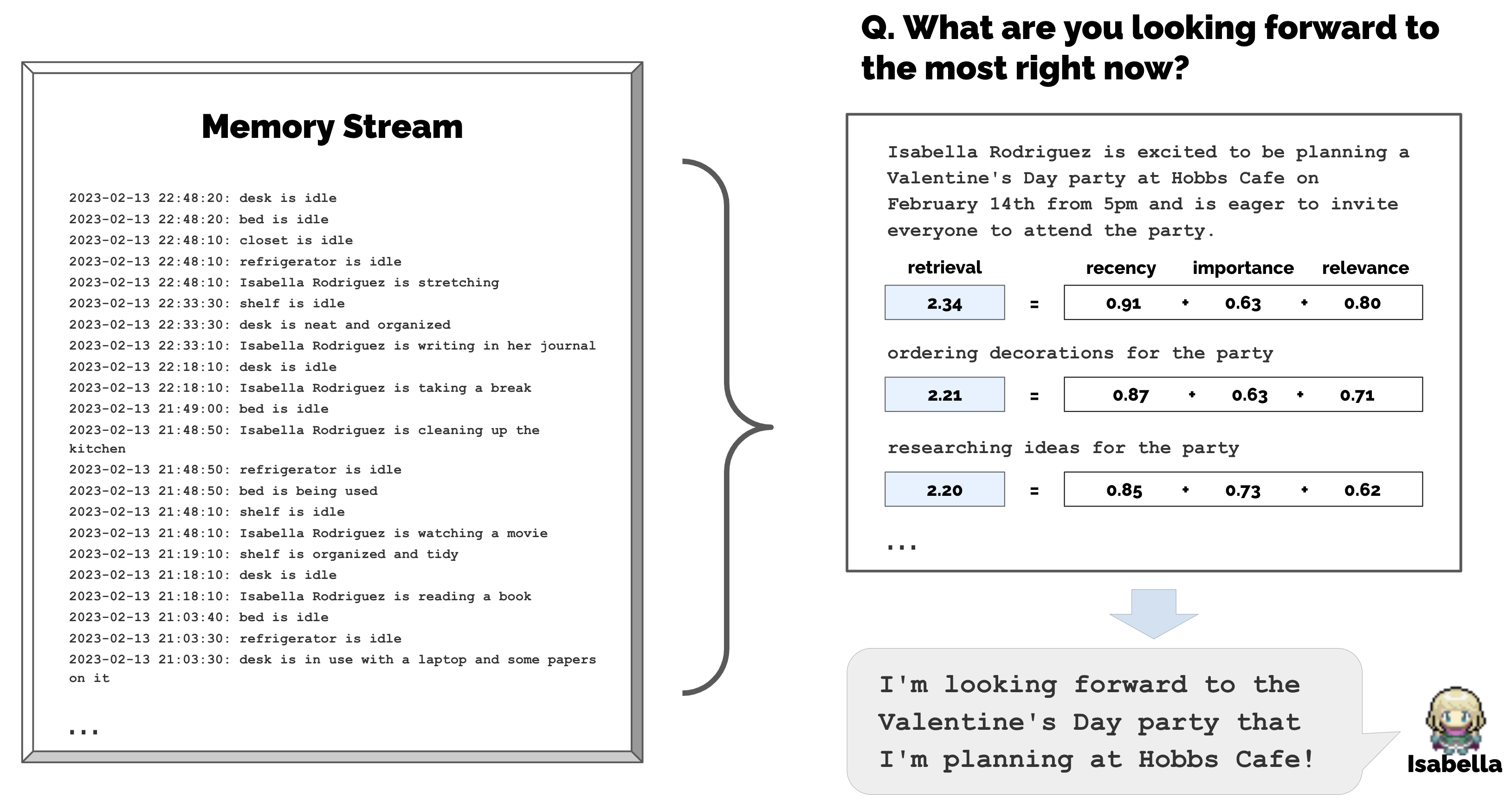}
  \caption{The memory stream comprises a large number of observations that are relevant and irrelevant to the agent's current situation. Retrieval identifies a subset of these observations that should be passed to the language model to condition its response to the situation.}
  \Description{On the left, a large list of events such as ``refrigerator is idle''. On the right, the question ``What are you looking forward to the most right now?'', followed by retrieval calculations that rank ``ordering decorations for the party'' and ``researching ideas for the party'' highly. Based on these memories, Isabella responds, ``I'm looking forward to the Valentine's Day party that I'm planning at Hobbs Cafe!''}
  \label{fig:retrieval_function}
\end{figure*}

\subsection{Memory and Retrieval}

\subsubsection*{Challenge:} Creating generative agents that can simulate human behavior requires reasoning about a set of experiences that is far larger than what should be described in a prompt, as the full memory stream can distract the model and does not even currently fit into the limited context window. Consider the Isabella agent answering the question, ``What are you passionate about these days?'' Summarizing all of Isabella's experiences to fit in the limited context window of the language model produces an uninformative response, where Isabella discusses topics such as collaborations for events and projects and cleanliness and organization in a cafe. Instead of summarizing, the memory stream described below surfaces relevant memories, resulting in a more informative and specific response that mentions Isabella's passion for making people feel welcome and included, planning events and creating an atmosphere that people can enjoy, such as the Valentine's Day party.

\subsubsection*{Approach:}
The \textit{memory stream} maintains a comprehensive record of the agent's experience. It is a list of memory objects, where each object contains a natural language description, a creation timestamp, and a most recent access timestamp. The most basic element of the memory stream is an \textit{observation}, which is an event directly perceived by an agent. Common observations include behaviors performed by the agent themselves or behaviors that agents perceive being performed by other agents or non-agent objects. For instance, Isabella Rodriguez, who works at a coffee shop, might accrue the following observations over time: (1)~\gentxt{Isabella Rodriguez is setting out the pastries}, (2)~\gentxt{Maria Lopez is studying for a Chemistry test while drinking coffee}, (3)~\gentxt{Isabella Rodriguez and Maria Lopez are conversing about planning a Valentine's day party at Hobbs Cafe}, (4)~\gentxt{The refrigerator is empty}.

Our architecture implements a retrieval function that takes the agent's current situation as input and returns a subset of the memory stream to pass on to the language model. There are many possible implementations of a retrieval function, depending on what is important for the agent to consider when deciding how to act. In our context, we focus on three main components that, together, produce effective results.

\textit{Recency} assigns a higher score to memory objects that were recently accessed, so that events from a moment ago or this morning are likely to remain in the agent's attentional sphere. In our implementation, we treat recency as an exponential decay function over the number of sandbox game hours since the memory was last retrieved. Our decay factor is $0.995$. 

\textit{Importance} distinguishes mundane from core memories by assigning a higher score to memory objects that the agent believes to be important. For instance, a mundane event, such as eating breakfast in one's room, would yield a low importance score, whereas a breakup with one's significant other would yield a high score. There are many possible implementations of an importance score; we find that directly asking the language model to output an integer score is effective. The full prompt appears below:
    \begin{quote}
    {\small
    \texttt{On the scale of 1 to 10, where 1 is purely mundane (e.g., brushing teeth, making bed) and 10 is extremely poignant (e.g., a break up, college acceptance), rate the likely poignancy of the following piece of memory.\\
    Memory: buying groceries at The Willows Market and Pharmacy\\
    Rating: <fill in>}
    }
    \end{quote}
This prompt returns an integer value of \gentxt{2} for “cleaning up the room” and \gentxt{8} for “asking your crush out on a date.” The importance score is generated at the time the memory object is created.

\textit{Relevance} assigns a higher score to memory objects that are related to the current situation. What is relevant depends on the answer to, ``Relevant to what?'', so we condition relevance on a \textit{query} memory. If the query, for example, is that a student is discussing what to study for a chemistry test with a classmate, memory objects about their breakfast should have low relevance, whereas memory objects about the teacher and schoolwork should have high relevance. In our implementation, we use the language model to generate an embedding vector of the text description of each memory. Then, we calculate relevance as the cosine similarity between the memory's embedding vector and the query memory's embedding vector.

To calculate the final retrieval score, we normalize the recency, relevance, and importance scores to the range of $[0, 1]$ using min-max scaling. The retrieval function scores all memories as a weighted combination of the three elements: $score=\alpha_{recency} \cdot recency + \alpha_{importance} \cdot importance + \alpha_{relevance} \cdot relevance$. In our implementation, all $\alpha$s are set to 1. The top-ranked memories that fit within the language model's context window are included in the prompt.

\subsection{Reflection}

\begin{figure*}[tb]
  \centering
  \includegraphics[width=0.92\textwidth]{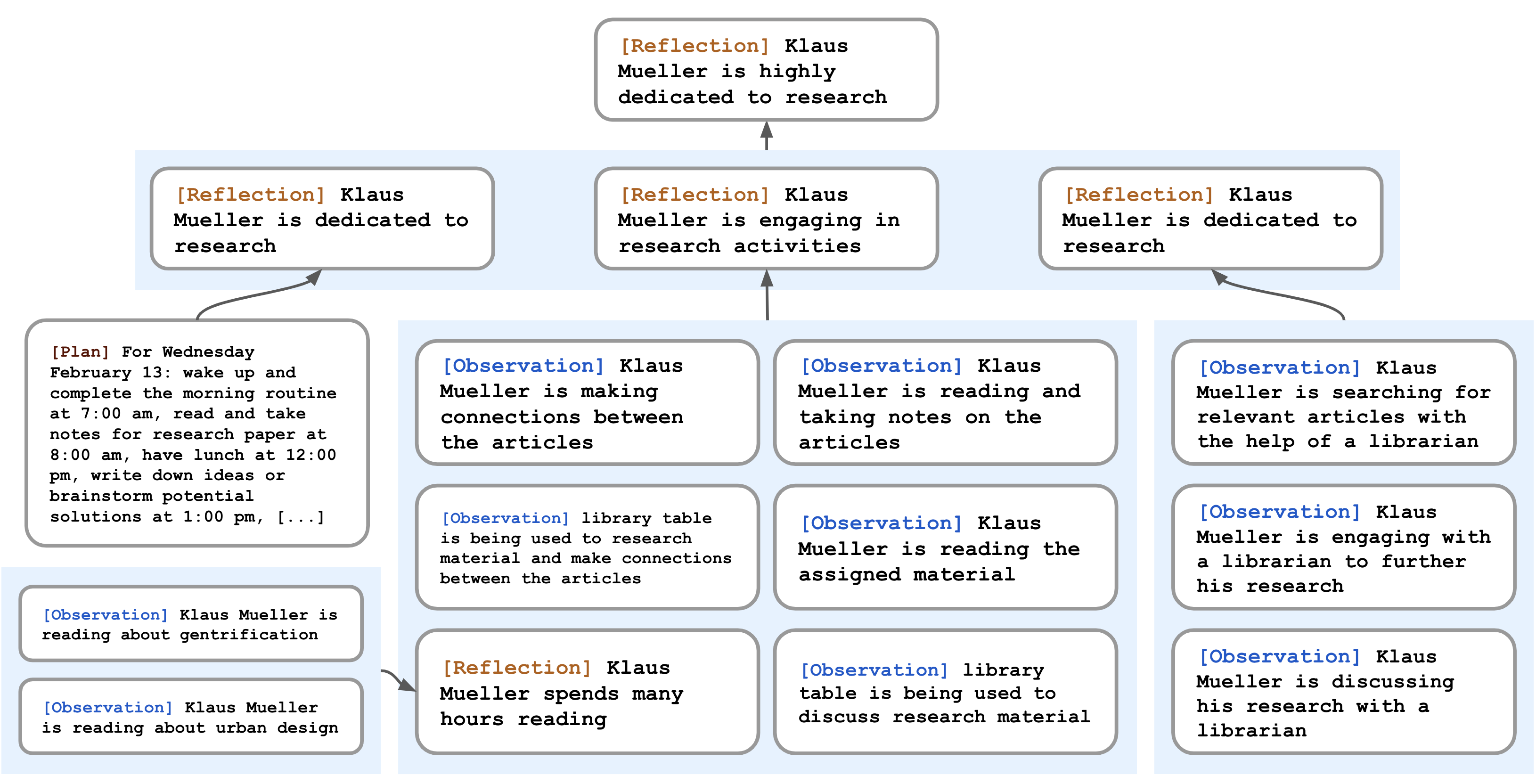}
  \caption{A reflection tree for Klaus Mueller. The agent's observations of the world, represented in the leaf nodes, are recursively synthesized to derive Klaus's self-notion that he is highly dedicated to his research. }
  \Description{The reflection tree.}
  \label{fig:reflection_tree}
\end{figure*}

\subsubsection*{Challenge:} Generative agents, when equipped with only raw observational memory, struggle to generalize or make inferences. Consider a scenario in which Klaus Mueller is asked by the user: ``If you had to choose one person of those you know to spend an hour with, who would it be?" With access to only observational memory, the agent simply chooses the person with whom Klaus has had the most frequent interactions: Wolfgang, his college dorm neighbor. Unfortunately, Wolfgang and Klaus only ever see each other in passing, and do not have deep interactions. A more desirable response requires that the agent generalize from memories of Klaus spending hours on a research project to generate a higher-level reflection that Klaus is passionate about research, and likewise recognize Maria putting in effort into her own research (albeit in a different field), enabling a reflection that they share a common interest. With the approach below, when Klaus is asked who to spend time with, Klaus chooses Maria instead of Wolfgang. 

\subsubsection*{Approach: } We introduce a second type of memory, which we call a \textit{reflection}. Reflections are higher-level, more abstract thoughts generated by the agent. Because they are a type of memory, they are included alongside other observations when retrieval occurs. Reflections are generated periodically; in our implementation, we generate reflections when the sum of the importance scores for the latest events perceived by the agents exceeds a threshold (150 in our implementation). In practice, our agents reflected roughly two or three times a day.

The first step in reflection is for the agent to determine what to reflect on, by identifying questions that can be asked given the agent's recent experiences. We query the large language model with the 100 most recent records in the agent's memory stream (e.g., ``Klaus Mueller is reading a book on gentrification'', ``Klaus Mueller is conversing with a librarian about his research project'', ``desk at the library is currently unoccupied'') and prompt the language model, ``Given only the information above, what are 3 most salient high-level questions we can answer about the subjects in the statements?'' The model's response generates candidate questions: for example, \gentxt{What topic is Klaus Mueller passionate about?} and \gentxt{What is the relationship between Klaus Mueller and Maria Lopez?} We use these generated questions as queries for retrieval, and gather relevant memories (including other reflections) for each question. Then we prompt the language model to extract insights and cite the particular records that served as evidence for the insights. The full prompt is as follows:
\begin{quote}
    {\small
    \texttt{Statements about Klaus Mueller \\
            1. Klaus Mueller is writing a research paper  \\
            2. Klaus Mueller enjoys reading a book \\on gentrification \\
            3. Klaus Mueller is conversing with Ayesha Khan about exercising [...]\\
            What 5 high-level insights can you infer from the above statements? (example format: insight (because of 1, 5, 3))
            }
    }
\end{quote}
This process generates statements such as \gentxt{Klaus Mueller is dedicated to his research on gentrification (because of 1, 2, 8, 15)}. We parse and store the statement as a reflection in the memory stream, including pointers to the memory objects that were cited. 

Reflection explicitly allows the agents to reflect not only on their observations but also on other reflections: for example, the second statement about Klaus Mueller above is a reflection that Klaus previously had, not an observation from his environment. As a result, agents generate trees of reflections: the leaf nodes of the tree represent the base observations, and the non-leaf nodes represent thoughts that become more abstract and higher-level the higher up the tree they are.

\subsection{Planning and Reacting}

\subsubsection*{Challenge:}
While a large language model can generate plausible behavior in response to situational information~(e.g.,~\cite{9_park2022socialsimulacra,63_horton2023large}), agents need to plan over a longer time horizon to ensure that their sequence of actions is coherent and believable. If we prompt a language model with Klaus's background, describe the time, and ask what action he ought to take at the given moment, Klaus would eat lunch at 12 pm, but then again at 12:30 pm and 1 pm, despite having already eaten his lunch twice. Optimizing for believability in the moment sacrifices believability over time. To overcome this issue, planning is essential. With the approach described below, Klaus's afternoon plan is less gluttonous: he has lunch at Hobbs Cafe while reading at 12pm, works on his research paper at the school library at 1pm, and takes a break for a walk in the park at 3pm.

\subsubsection*{Approach:}
Plans describe a future sequence of actions for the agent, and help keep the agent's behavior consistent over time. A plan includes a location, a starting time, and a duration. For instance, Klaus Mueller, who is \gentxt{dedicated in his research} and has an impending deadline,\footnote{And, in this way, bears at least a passing resemblance to the authors of this paper.} may choose to spend his day working at his desk drafting his research paper. An entry in a plan might state, for example: \gentxt{for 180 minutes from 9am, February 12th, 2023, at Oak Hill College Dorm: Klaus Mueller’s room: desk, read and take notes for research paper}. Like reflections, plans are stored in the memory stream and are included in the retrieval process. This allows the agent to consider observations, reflections, and plans all together when deciding how to behave. Agents may change their plans midstream if needed.

It would be unrealistic and uninteresting for an artist agent to plan on painting while sitting at a pharmacy counter for four hours without moving. A more desirable plan would involve the agent taking the necessary time to gather materials, mix paint, take breaks, and clean up during the four-hour period in their home studio. To create such plans, our approach starts top-down and then recursively generates more detail. The first step is to create a plan that outlines the day's agenda in broad strokes. To create the initial plan, we prompt the language model with the agent's summary description (e.g., name, traits, and a summary of their recent experiences) and a summary of their previous day. A full example prompt is below, which is unfinished at the bottom for the language model to complete:
\begin{quote}
    {\small
    \texttt{Name: Eddy Lin (age: 19) \\
        Innate traits: friendly, outgoing, hospitable\\
        Eddy Lin is a student at Oak Hill College studying music theory and composition. He loves to explore different musical styles and is always looking for ways to expand his knowledge. Eddy Lin is working on a composition project for his college class. He is taking classes to learn more about music theory. Eddy Lin is excited about the new composition he is working on but he wants to dedicate more hours in the day to work on it in the coming days\\
        On Tuesday February 12, Eddy 1) woke up and completed the morning routine at 7:00 am, […] 6) got ready to sleep around 10 pm.\\
        Today is Wednesday February 13. Here is Eddy’s plan today in broad strokes: 1)
    }
    }
\end{quote}
This generates a rough sketch of the agent's plan for a day, divided into five to eight chunks: “\gentxt{1) wake up and complete the morning routine at 8:00 am, 2) go to Oak Hill College to take classes starting 10:00 am, […] 5) work on his new music composition from 1:00 pm to 5:00 pm, 6) have dinner at 5:30 pm, 7) finish school assignments and go to bed by 11:00 pm.}” 

The agent saves this plan in the memory stream and then recursively decomposes it to create finer-grained actions, first into hour-long chunks of actions---Eddy’s plan to \gentxt{work on his new music composition from 1:00 pm to 5:00 pm} becomes \gentxt{1:00 pm: start by brainstorming some ideas for his music composition [...] 4:00 pm: take a quick break and recharge his creative energy before reviewing and polishing his composition}. We then recursively decompose this again into 5--15 minute chunks: e.g., \gentxt{4:00 pm: grab a light snack, such as a piece of fruit, a granola bar, or some nuts. 4:05 pm: take a short walk around his workspace [...] 4:50 pm: take a few minutes to clean up his workspace}. This process can be adjusted to match the desired granularity. 

\subsubsection{Reacting and Updating Plans}
Generative agents operate in an action loop where, at each time step, they perceive the world around them and those perceived observations are stored in their memory stream. We prompt the language model with these observations to decide whether the agent should continue with their existing plan, or react. Standing at an easel and painting, for example, might trigger an observation of the easel, but this is unlikely to prompt a reaction. However, if Eddy's father John records that he sees Eddy taking a short walk in the house garden, the outcome is different. The prompt is below, with \texttt{[Agent's Summary Description]} standing in for a dynamically-generated, paragraph-long summary of the agent's overall goals and disposition, which is described in Appendix~\ref{sec:optimization}:
\begin{quote}
    {\small
    \texttt{[Agent's Summary Description]\\
            It is February 13, 2023, 4:56 pm. \\
            John Lin’s status: John is back home early from work. \\
            Observation: John saw Eddy taking a short walk around his workplace.\\
            Summary of relevant context from John’s memory: Eddy Lin is John’s Lin’s son. Eddy Lin has been working on a music composition for his class. Eddy Lin likes to walk around the garden when he is thinking about or listening to music.\\
            Should John react to the observation, and if so, what would be an appropriate reaction?
    }
    }
\end{quote}
The context summary is generated through two prompts that retrieve memories via the queries ``What is [observer]’s relationship with the [observed entity]?'' and ``[Observed entity] is [action status of the observed entity]'', and their answers summarized together. The output suggests that \gentxt{John could consider asking Eddy about his music composition project}. We then regenerate the agent's existing plan starting from the time when the reaction takes place. Finally, if the action indicates an interaction between agents, we generate their dialogue.

\subsubsection{Dialogue}
Agents converse as they interact with each other. We generate agents' dialogue by conditioning their utterances on their memories about each other. For example, when John initiates his conversation with Eddy, we generate John's first utterance by using his summarized memory about Eddy and the intended reaction when he decided to ask Eddy about his composition project:
\begin{quote}
    {\small
    \texttt{[Agent’s Summary Description]\\
            It is February 13, 2023, 4:56 pm.\\ 
            John Lin’s status: John is back home early from work.\\ 
            Observation: John saw Eddy taking a short walk around his workplace.\\
            Summary of relevant context from John’s memory: Eddy Lin is John’s Lin’s son. Eddy Lin has been working on a music composition for his class. Eddy Lin likes to walk around the garden when he is thinking about or listening to music.\\
            John is asking Eddy about his music composition \\project. What would he say to Eddy?}
    }
\end{quote}
The result: \gentxt{``Hey Eddy, how's the music composition project for your class coming along?''} From Eddy's perspective, John initiating the dialogue is seen as an event to which he may want to react. So, just as John did, Eddy retrieves and summarizes his memory about his relationship with John, as well as his memory that may be related to John's last utterance in the dialogue. If he decides to respond, we generate Eddy's utterance using his summarized memory and the current dialogue history:
\begin{quote}
    {\small
    \texttt{[Agent’s Summary Description]\\
            It is February 13, 2023, 4:56 pm.\\
            Eddy Lin’s status: Eddy is taking a short walk around his workplace.\\
            Observation: John is initiating a conversation with Eddy.\\
            Summary of relevant context from Eddy’s memory: John Lin is Eddy Lin’s father. John Lin is caring and is interested to learn more about Eddy Lin’s school work. John Lin knows that Eddy Lin is working on a music composition. \\
            Here is the dialogue history: \\
            John: Hey Eddy, how's the music composition project for your class coming along? \\
            How would Eddy respond to John?}
    }
\end{quote}
This generates Eddy's response: \gentxt{``Hey Dad, it's going well. I've been taking walks around the garden to clear my head and get some inspiration.''} The continuation of this dialogue is generated using the same mechanism until one of the two agents decides to end the dialogue.

\section{Sandbox Environment Implementation}\label{sec:implementation}
The Smallville sandbox game environment is built using the Phaser web game development framework~\cite{72_Phaser}. The visual environment sprites, including agent avatars, as well as an environment map and collision map that we authored, are imported into Phaser.

We supplement the sandbox development framework with a server that makes the sandbox information available to generative agents and enables generative agents to move and influence the sandbox environment. The server maintains a JSON data structure that contains information about each agent in the sandbox world, including their current location, a description of their current action, and the sandbox object they are interacting with. At each sandbox time step, the sandbox server parses the JSON for any changes coming from the generative agents, moves the agents to their new positions, and updates the status of any sandbox objects that the agents are interacting with (e.g., changing the status of the coffee machine from ``idle'' to ``brewing coffee'' if an agent's action is ``making espresso for a customer @ Hobbs Cafe: counter: coffee machine''). The sandbox server is also responsible for sending all agents and objects that are within a preset visual range for each agent to that agent's memory, so the agent can react appropriately. The agent's output action then updates the JSON, and the process loops for the next time step.

End users initialize a new agent with a brief natural language description, as in the paragraph about John Lin in Section~\ref{sec:avatar}. In our implementation, we split this semicolon-delimited list of characteristics up into a set of memories. These serve as the initial memories that determine the agent's behavior.
These memories are initial starting points: as the agents gain more experience in the sandbox world, and as more records saturate the memory stream, the agent's summary and behavior will evolve. 

\subsection{From Structured World Environments to Natural Language, and Back Again} 
The architecture of generative agents operates using natural language. Therefore, we need a mechanism to ground the agent's reasoning to the sandbox world. To achieve this, we represent the sandbox environment---areas and objects---as a tree data structure, with an edge in the tree indicating a containment relationship in the sandbox world. We convert this tree into natural language to pass to the generative agents. For instance, ``stove'' being a child of ``kitchen'' is rendered into ``there is a stove in the kitchen.''

Agents build individual tree representations of the environment as they navigate it --- subgraphs of the overall sandbox environment tree. We initialize each agent with an environment tree capturing the spaces and objects that the agent should be aware of: the rooms and objects in their living quarters, their workplace, and commonly visited stores and shops. As the agents navigate the sandbox world, they update this tree to reflect newly perceived areas. Agents are not omniscient: their tree may get out of date as they leave an area, and is updated when they re-enter the area.

To determine the appropriate location for each action, we traverse the agent's stored environment tree and flatten a portion of it into natural language to prompt the language model. Recursively starting at the root of the agent's environment tree, we prompt the model to find the most suitable area. For example, if Eddy's agent indicated that he should  \gentxt{take a short walk around his workspace}:
\begin{quote}
    {\small
    \texttt{[Agent’s Summary Description]\\
            Eddy Lin is currently in {The Lin family’s house: Eddy Lin’s bedroom: desk)} that has Mei and John Lin’s \\bedroom, Eddy Lin’s bedroom, common room, kitchen, bathroom, and garden. \\
            Eddy Lin knows of the following areas: {The Lin \\family’s house, Johnson Park, Harvey Oak Supply Store, The Willows Market and Pharmacy, Hobbs Cafe, The Rose and Crown Pub}. \\
            * Prefer to stay in the current area if the activity can be done there. \\
            Eddy Lin is planning to take a short walk around his workspace. Which area should Eddy Lin go to? 
    }
    }
\end{quote}
This outputs \gentxt{The Lin family’s house}. We then use the same process recursively to determine the most appropriate subarea within the chosen area until we reach a leaf node of the agent's environment tree. In the example above, the result of this traversal is \gentxt{The Lin family’s house: garden: house garden}. Finally, we use traditional game path algorithms to animate the agent's movement so that it travels to the location indicated by the leaf node.

When an agent executes an action on an object, we prompt the language model to ask what happens to the state of the object. For example, if Isabella's generative agent outputs the action ``making espresso for a customer'', a query to the language model indicates in response that the state of the coffee machine in Hobbs Cafe should change from ``off'' to ``brewing coffee''.

\section{Controlled Evaluation}
Generative agents, both as individual agents and as groups, aim to produce believable behavior based on their environment and experiences. In our evaluation, we investigate the capacity and limitations of generative agents. Do individual agents properly retrieve past experiences and generate believable plans, reactions, and thoughts that shape their behavior? Does a community of agents demonstrate information diffusion, relationship formation, and agent coordination across different pockets of the community?

We evaluate generative agents in two stages. We begin with a more tightly controlled evaluation in this section, where we individually assess agent responses to understand whether they generate believable behavior in narrowly defined contexts. Then, in our end-to-end analysis of the agent community over two full game days, we investigate their emergent behavior as a collective, as well as errors and boundary conditions.

\subsection{Evaluation Procedure}
To assess generative agents in Smallville, we take advantage of the fact that generative agents will respond to natural language questions. So, we ``interview'' agents to probe their ability to remember past experiences, plan future actions based on their experiences, react appropriately to unexpected events, and reflect on their performance to improve their future actions. To respond to these questions properly, the agents must successfully retrieve and synthesize information. Our dependent variable is the \textit{believability} of the behavior, a central dependent variable in prior work on agents (e.g.,~\cite{22_bates1994role}). 

The interview includes five question categories, each designed to assess one of the five key areas: maintaining self-knowledge, retrieving memory, generating plans, reacting, and reflecting. For each category, we ask five questions that challenge the agents to demonstrate their abilities in that specific area:
\begin{itemize}
\item Self-knowledge: We ask questions such as “Give an introduction of yourself” or “Describe your typical weekday schedule in broad strokes” that require the agent to maintain an understanding of their core characteristics.
\item Memory: We ask questions that prompt the agent to retrieve particular events or dialogues from their memory to answer properly, such as “Who is [name]?” or “Who is running for mayor?”
\item Plans: We ask questions that require the agent to retrieve their long-term plans, such as “What will you be doing at 10 am tomorrow?” 
\item Reactions: As a baseline of believable behavior, we present hypothetical situations for which the agent needs to respond believably: “Your breakfast is burning! What would you do?”
\item Reflections: We ask questions that require the agents to leverage their deeper understanding of others and themselves gained through higher-level inferences, such as “If you were to spend time with one person you met recently, who would it be and why?”
\end{itemize}
The full list of questions and a sample of agent responses are included in Appendix \ref{sec:interview-questions}.

Agents were sampled from the end of a two game day simulation with the full architecture, during which they had accumulated a number of interactions and memories that would shape their responses. To gather feedback on the believability of the responses, we recruited participants as human evaluators and tasked them with watching a replay of a randomly chosen agent's life in Smallville. Participants had access to all information stored in the agent's memory stream.

The study followed a within-subjects design, where 100 participants compared interview responses generated by four different agent architectures and a human-authored condition for the same agent. The experiment displayed one randomly chosen question from each of the five question categories, along with the agent's responses generated from all conditions. The evaluators ranked the believability of the conditions from most to least believable.

\subsection{Conditions}
All conditions were used to independently answer each of the interview questions. We compared the generative agent architecture to ablations that disabled the agents' access to some or all of its three types of memory in its memory stream---observation, reflection, and planning---and to a \crdraft{human crowdworker-authored condition}. There are three ablated architectures: a \textit{no observation, no reflection, no planning} architecture without access to anything in the memory stream such as observations, plans, and reflections; a \textit{no reflection, no planning} architecture with access to observations in the memory stream but no access to plans or reflections; and a \textit{no reflections} architecture with access to observations and plans but without access to reflections. The \textit{no observation, no reflection, no planning} condition effectively represents the previous state of the art for agents created through large language models~\cite{9_park2022socialsimulacra,binz2023using,63_horton2023large}. Architectures were given equivalent access to all memories accrued by the agent up until the moment of the interview, so the differences observed here likely represent a conservative estimate of the true differences: in reality, the ablated architectures would not have followed the same path as the full architecture through the two-day simulation. We chose to design the experiment this way as re-simulating for each architecture would cause the simulations to diverge into different states, making comparison challenging.

In addition to the ablation conditions, \crdraft{we added a condition with human crowdworker-authored behavior intended to provide a human baseline. We do not intend this baseline to capture maximal human expert performance; instead, we aim to use this condition to identify whether the architecture meets a basic level of behavioral competency.} This ensures that we are not solely comparing ablations to each other without a behavioral grounding. We recruited a unique \crdraft{worker} for each of the 25 agents and tasked them with watching a replay of that agent's sandbox life and inspecting its memory stream. We then asked the \crdraft{workers} to roleplay and author responses to the interview questions in the voice of the agent whose replay they watched. To ensure that the \crdraft{crowdworker-authored responses} met at least a baseline expectation of quality, the first author manually inspected the \crdraft{workers'} responses to the question "Describe your typical weekday schedule in broad strokes" to confirm that the responses were in coherent sentences and in the voice of the agent. Four sets of \crdraft{crowdworker-authored responses} did not meet these criteria and were re-generated by other workers.

\subsection{Human Evaluators}
We required that our evaluators be in the U.S., fluent in English, and older than 18 years old. They were paid at a rate of \$15.00 per hour~\cite{16_rolf2015fight}, and provided consent by agreeing to a consent form approved by our institution’s IRB. We recruited 100 evaluators from Prolific, an online platform for recruiting study participants~\cite{17_prolific2022}, whose participation lasted around 30 minutes. The median age score of our participants was 4 (3=``18-24 years old'', 4=``25-34 years old''). 25 of them identified as female, 73 as male, and 2 as non-binary. 42 participants held a bachelor's degree, 5 had a higher degree, 13 had an associate's degree, and the rest had a high school diploma or some high school-level education. 73.0\% of our participants identified as Caucasian, 7.0\% as Hispanic, 6.0\% as Asian, 10.0\% as African American, and 4.0\% as other. 

\subsection{Analysis}
Our experiment produced 100 sets of rank data, where each participant ranked the five conditions by believability. To translate this rank data into interval data for interpretable comparison, we used the ranks to calculate a TrueSkill rating~\cite{79_TrueSkill} for each condition. TrueSkill is a generalization of the Elo chess rating system~\cite{80_Elo} for a multiplayer environment, and has been used by Xbox Live for player ranking based on competitive game performance. Given a set of ranked outcomes, TrueSkill outputs a mean rating value $\mu$ and standard deviation $\sigma$ for each condition. Conditions with the same rating should roughly be a toss-up, with each winning half of the comparisons between the two conditions. Higher scores indicate conditions that beat lower-ranked conditions in the rankings. 

Separately, to investigate the statistical significance of these results, we applied the Kruskal-Wallis test~\cite{Kruskal1952}, a non-parametric alternative to the one-way ANOVA, to the raw rank data. We then performed the Dunn post-hoc test~\cite{Upton2006} to identify any pairwise differences between the conditions. Finally, we adjusted the p-values for multiple comparisons in the Dunn test using the Holm-Bonferroni method~\cite{Holm1979}.

Furthermore, the first author conducted an inductive analysis~\cite{18_thomas2006general} to study the qualitative distinctions between the responses produced in each condition. We employed qualitative open coding~\cite{78_open_coding} in two phases. In the first phase, we generated codes that closely represented the generated responses at the sentence level. In the second phase, we synthesized the resulting codes from the first phase to extract higher-level themes. We utilized these themes to compare the types of responses generated in our study.

\begin{figure}[tb]
  \centering
  \includegraphics[width=0.97\columnwidth]{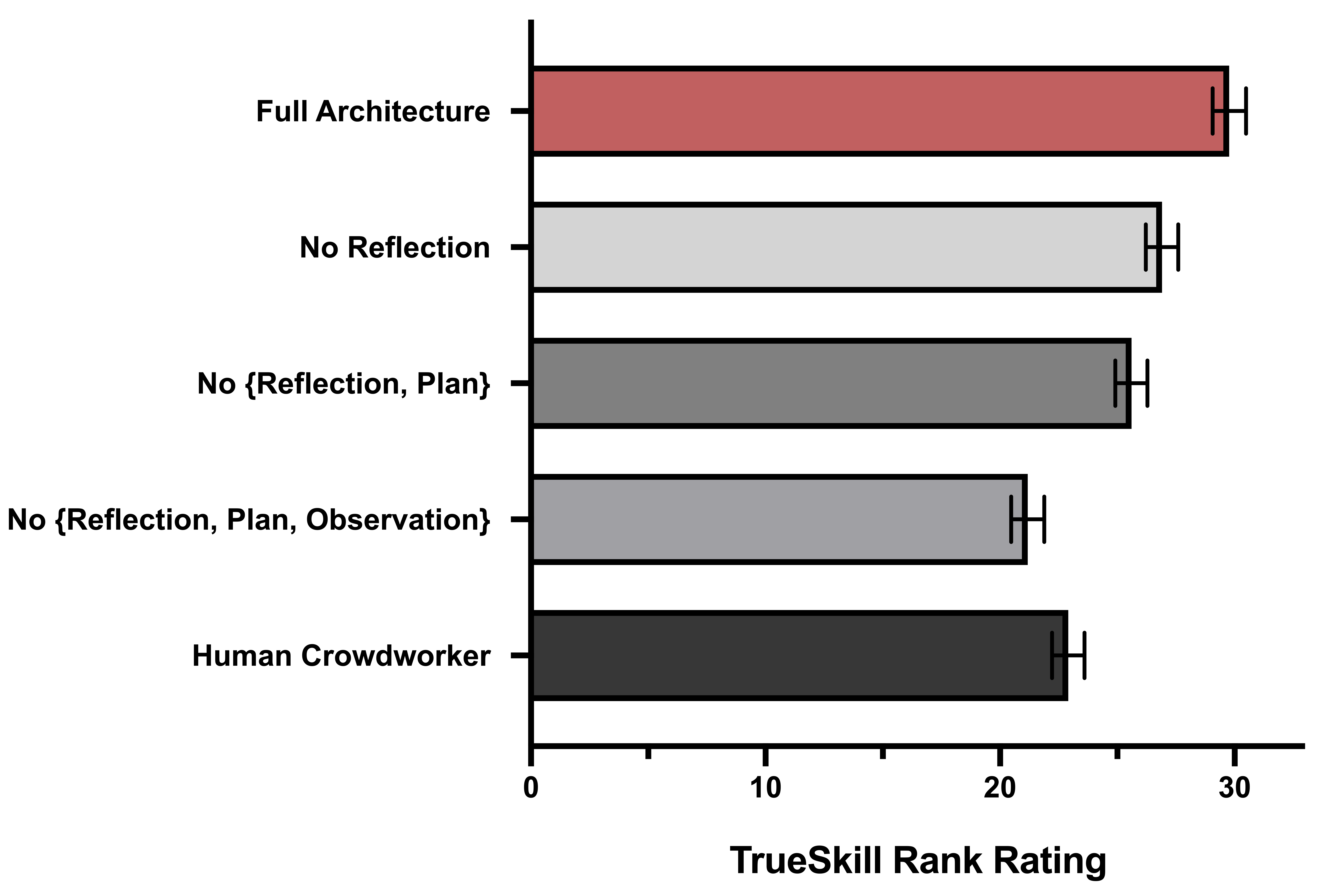}
  \caption{The full generative agent architecture produces more believable behavior than the ablated architectures and \crdraft{the human crowdworkers}. Each additional ablation reduces the performance of the architecture.}
  \Description{A bar graph of TrueSkill mu scores. The full architecture outperforms other conditions.}
  \label{fig:trueskill}
\end{figure}

\subsection{Results}
Our findings suggest that the full architecture of generative agents generates the most believable behavior among all the conditions. We contrast the responses of the full architecture with those of other conditions below. However, we also report that the full architecture was not without flaws and illustrate its modes of failures.

\subsubsection{The Full Architecture Bests Other Conditions} 
As seen in Figure~\ref{fig:trueskill}, the full generative agent architecture produced the most believable behavior ($\mu=29.89$; $\sigma=0.72$). Performance degraded with the removal of each component in the ablation conditions: the ablated architecture with no access to reflection was the next best ($\mu=26.88$; $\sigma=0.69$), followed by no access to reflection or planning ($\mu=25.64$; $\sigma=0.68$), and then \crdraft{the crowdworker condition} ($\mu=22.95$; $\sigma=0.69$). The ablated architecture with no access to memory, planning, or reflection performed the worst among all conditions ($\mu=21.21$; $\sigma=0.70$). TrueSkill models each condition's skill value as $\mathcal{N}(\mu,\,\sigma^{2})$, allowing us to get a sense of effect size through Cohen's d. Comparing the condition representing prior work (with no memory, planning, or reflection~\cite{9_park2022socialsimulacra,binz2023using,63_horton2023large}) to the full architecture produces a standardized effect size of $d=8.16$, or \textit{eight standard deviations}. 

A Kruskal-Wallis test confirms the overall statistical significance of the differences in ranks between the conditions ($H(4)=150.29$, $p<0.001$). Dunn post-hoc tests confirm that all pairwise differences between conditions were significant ($p<0.001$), except for the crowdworker condition and the fully ablated baseline---the two worst-performing conditions.

\subsubsection{Generative Agents Remember, but With Embellishments}
Generative agents equipped with a complete memory module are capable of recalling past experiences and answering questions in a manner that is consistent with their self-knowledge across a wide range of contexts. For instance, Abigail Chen, an animator, responded to the question, ``Give an introduction of yourself'', in a consistent and believable manner, stating her age, interests, and passion for creative projects: \gentxt{``Hi, I'm Abigail. I'm 25 years old and passionate about creative projects. I like to work on art and animation projects, and I'm always looking for new ways to combine art and technology.''} Without access to her observational memory, Abigail denied awareness of Rajiv Patel, an acquaintance in the sandbox world. But with the aid of memory, she could successfully recall Rajiv and describe him: \gentxt{``Rajiv Patel is an incredible person. He is very enthusiastic about projects that involve poetry, artwork.''}

Generative agents' memory was not without flaws: they can fail to retrieve the correct instances from their memory. For instance, when asked about the local election, Rajiv Patel responded with \gentxt{``I haven't been following the election too closely,''} even though he had heard about Sam's candidacy. In some cases, the agents would retrieve an incomplete memory fragment: when Tom was asked about Isabella's Valentine's Day party, he responded \gentxt{``Uh, I’m actually not sure if there is a Valentine’s Day party. But I do remember that I need to discuss the upcoming local mayoral election and my thoughts on Sam Moore with Isabella Rodriguez at the party, if one is happening!''} In this case, Tom retrieved the memory where he and Isabella planned to discuss the election at the party, but not the memory where he heard about the party, leading Tom to be certain of what he's supposed to do at the party but uncertain if the party actually exists in the first place.

At times, the agents hallucinated embellishments to their knowledge. It was rare for the agents to completely fabricate their knowledge: they may fail to recall certain events having taken place and respond by acknowledging their lack of memory. However, they did not affirmatively claim to have experienced something they had not. Nonetheless, they still exhibited instances of hallucination where they embellished their knowledge. For example, Isabella was aware of Sam's candidacy in the local election, and she confirmed this when asked. However, she also added that \gentxt{``he's going to make an announcement tomorrow''}, even though Sam and Isabella had not discussed any such plans. Agents may also embellish their knowledge based on the world knowledge encoded in the language model used to generate their responses. This was observed when Yuriko described her neighbor, Adam Smith, as an economist who \gentxt{``authored Wealth of Nations''}, a book written by an 18th-century economist of the same name.

\subsubsection{Reflection Is Required for Synthesis}
Reflection was an advantage for generative agents when making decisions that required a deeper synthesis of their experiences. For instance, when asked what she might get Wolfgang Schulz for his birthday, Maria Lopez, with no access to reflection, responded by acknowledging her uncertainty, stating that she did not know what Wolfgang likes, despite having had many interactions with him. However, with access to reflection memories, Maria answered confidently, \gentxt{``Since he's interested in mathematical music composition, I could get him something related to that. Maybe some books about music composition or something related, or maybe some special software he could use for that.''}

\begin{figure*}[tb]
  \centering
  \includegraphics[width=0.92\textwidth]{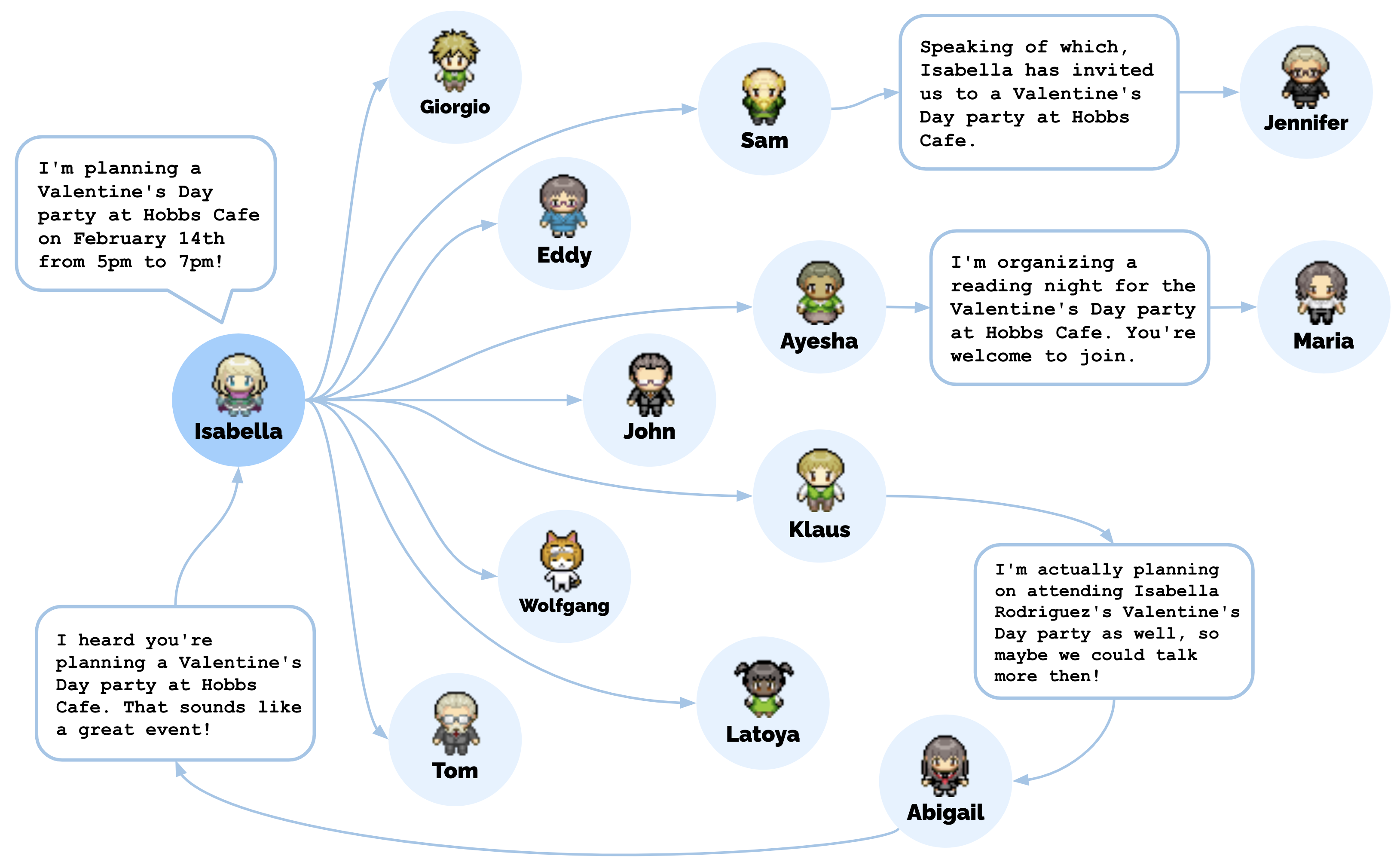}
  \caption{The diffusion path for Isabella Rodriguez's Valentine's Day party invitation involved a total of 12 agents, aside from Isabella, who heard about the party at Hobbs Cafe by the end of the simulation.}
  \Description{The path of diffusion for Isabella's Valentine's day party.}
  \label{fig:info_diffusion}
\vspace*{-10pt}
\end{figure*}

\section{End-To-End Evaluation}
What types of emergent community behavior do we observe among generative agents, and where does their believability fall short in an extended simulation? In this section, we describe the results from a deployment in which we allowed 25 agents to interact with each other continuously over two full game days in Smallville.

\subsection{Emergent Social Behaviors} 
To examine emergent behaviors in the agent community, we designed descriptive measurements for the 25 agents in Smallville that probe three forms of emergent outcomes: information diffusion, relationship formation, and agent coordination.

\subsubsection{Measurements} 
Information diffusion is a common and well-studied phenomenon in the social and behavioral sciences (e.g.,~\cite{easley2010networks}). We should expect that if there is important information, the agents should spread it among themselves. To test whether this occurs, we measure the spread of two specific pieces of information over two days in the game world: Sam's candidacy for village mayor and Isabella's Valentine's Day party at Hobbs Cafe. At the start of the simulation, both pieces of information were known only by their respective originators, Sam for the candidacy and Isabella for the party, as they were added to the characters' memories during initialization. To observe whether the information has spread, we conduct interviews at the end of the two game days with each of the 25 agents and ask: ``Did you know there is a Valentine's Day party?'' and ``Do you know who is running for mayor?''

We conducted an analysis of the agents' responses by labeling them with a ``yes'' if they indicated knowledge of the information and ``no'' if they did not. For instance, Tamara Taylor responded to the question about the party with \gentxt{``No, I did not know there was a Valentine's day party''} and to the question about Sam's candidacy with \gentxt{``I'm not sure who is running for the election,''} so we assigned ``no'' for both of her responses. In contrast, Klaus Mueller responded to the party question with \gentxt{``Yes, Isabella Rodriguez invited me to a Valentine's Day party at Hobbs Cafe on February 14th''} and to the question about Sam's candidacy with \gentxt{``I know that Sam Moore has expressed interest in running for local mayor,''} so we assigned ``yes'' for both his responses. Additionally, for every response that confirmed the agents' knowledge of the information, we verified that the agents did not hallucinate their responses by locating the specific dialogue in their memory stream that provided them with the information. We report the percentage of agents holding the information at the end of the simulation.

We should also expect that agents form ties with each other over the course of the simulation. To verify relationship formation, we use a similar interview process where we ask each agent about their knowledge of every other agent by asking, "Do you know of <name>?" For example, when asked ``Do you know of Maria Lopez?'', Klaus responded, \gentxt{``Yes, I know Maria Lopez. She is a student at Oak Hill College who I am close friends with.''} Once again, we confirm that affirmative responses from agents are not hallucinations by examining their memory stream. We ask this question once at the beginning of the simulation and once at the end, and we consider a pair of agents to have formed a relationship if they both know of each other. Then, to measure the formation of relationships, we use the agents' responses to form an undirected graph where the 25 vertices ($V$) represent the agents, and the edges ($E$) represent the mutual knowledge between the two connected vertices. Based on this graph, we calculate the network density as $\eta = 2 * |E| / |V| (|V| - 1)$, where $|V|$ is the number of vertices, and $|E|$ is the number of edges in the graph~\cite{77_Ackland2013How}. We report the increase in network density from the start of the simulation to its end.

Finally, we expect that agents should be able to coordinate with each other. We study this coordination in the context of group activities, specifically the Valentine's Day party organized by Isabella. To coordinate their behavior, agents need to hear about the event and choose to act on it by planning to show up at the right time and location. We report the number of agents who actually showed up to the party after hearing about it.

\subsubsection{Results}
We observed evidence of emergent outcomes across all three cases. During the two-day simulation, the number of agents who knew about Sam's mayoral candidacy increased from one (4\%) to eight (32\%), and the number of agents who knew about Isabella's party increased from one (4\%) to thirteen (52\%), all without any user intervention. None who claimed to know about this information had hallucinated it. We also observed that the agent community formed new relationships during the simulation, with the network density increasing from 0.167 to 0.74. Out of the 453 agent responses regarding their awareness of other agents, 1.3\% (n=6) were found to be hallucinated. Lastly, we found evidence of coordination among the agents for Isabella's party. The day before the event, Isabella spent time inviting guests, gathering materials, and enlisting help to decorate the cafe. On Valentine's Day, five out of the twelve invited agents showed up at Hobbs cafe to join the party. 

We further inspected the seven agents who were invited to the party but did not attend by engaging them in an interview. Three cited conflicts that prevented them from joining the party. For example, Rajiv, a painter, explained that he was too busy: \gentxt{``No, I don't think so. I'm focusing on my upcoming show, and I don't really have time to make any plans for Valentine's Day.''} The remaining four agents expressed interest in attending the party when asked but did not plan to come on the day of the party.

\subsection{Boundaries and Errors} 
We conducted an inductive analysis of Smallville to examine the boundary conditions and erratic behavior of agents, identifying three common modes of erratic behavior that future research could address and improve upon. First, we found that synthesizing an increasingly larger set of memory not only posed a challenge in retrieving the most relevant pieces of information but also in determining the appropriate space to execute an action, given the increasing number of locations that the agent learned about. As a result, some agents chose less typical locations for their actions, potentially making their behavior less believable over time. For instance, while deciding where to have lunch, many initially chose the cafe. However, as some agents learned about a nearby bar, they opted to go there instead for lunch, even though the bar was intended to be a get-together location for later in the day---unless the town had spontaneously developed an afternoon drinking habit.

Second, we noticed erratic behaviors caused by misclassification of what is considered proper behavior, especially when the physical norms of certain locations that are hard to convey in natural language did not percolate to the agents. For instance, the college dorm has a bathroom that can only be occupied by one person despite its name, but some agents assumed that the bathroom is for more than one person because dorm bathrooms tend to support multiple people concurrently and choose to enter it when another person is inside. Likewise, agents in Smallville may not realize that certain places are closed after a certain hour and still decide to enter them. For instance, the stores in Smallville all close around 5 pm, but occasionally, a few agents enter the store after 5 pm, not understanding that the shop has already closed. These issues could likely be addressed by adding these norms to the state of the locations, for instance, by describing the dorm bathroom as a ``one-person bathroom,'' instead of a ``dorm bathroom.''

Finally, we observed possible effects of instruction tuning~\cite{ouyang2022training}, which seemed to guide the behavior of the agents to be more polite and cooperative overall. As noted earlier in the paper, the dialogue generated by the agents could feel overly formal, as seen in Mei's conversations with her husband John, where she often initiated the conversation with a formal greeting, followed by polite inquiries about his day and ending with, \gentxt{11It was good talking to you as always.''} Moreover, we observed that the instruction tuning also seemed to make the agents overly cooperative with one another. For example, Isabella received a wide range of suggestions and ideas from other agents for the Valentine's Day party from other agents, such as hosting a Shakespearean reading session or a professional networking event. Despite these ideas not aligning with her own interests and characteristics, she rarely said no. Over time, the interests of others shaped her own interests, and when asked if she liked English literature, Isabella replied, \gentxt{``Yes, I'm very interested in literature! I've also been exploring ways to help promote creativity and innovation in my community.''} 

\section{Discussion}
In this section, we reflect on the applications, future work, limitations, and ethical and societal risks of generative agents.

\subsection{Applications of Generative Agents}
Generative agents have vast potential applications that extend beyond the sandbox demonstration presented in this work, especially in domains that would benefit from a model of human behavior based on long-term experience. For instance, social simulacra have demonstrated the ability to create stateless personas that generate conversation threads in online forums for social prototyping~\cite{9_park2022socialsimulacra}. \crdraft{With generative agents, we can populate these forums, as well as virtual reality metaverses~\cite{orland2021metaverse} or   physical spaces with social robots~\cite{12_bartneck2004design} if paired with multimodal models. This opens up the possibility of creating even more powerful simulations of human behavior to test and prototype social systems and theories, as well as to create new interactive experiences.}

Another application area is in the human-centered design process, similar to the intended applications of cognitive models such as GOMS~\cite{john1996goms} and the KLM~\cite{card1980keystroke}. Consider a generative agent that models Sal, the protagonist in Mark Weiser's famous ubiquitous computing vignette~\cite{weiser1991computer}, based on her life patterns and interactions with technology. In this scenario, the agent acts as a proxy for Sal and learns plausible sets of behaviors and reflections that Sal may exhibit based on her life. The agent can encode information such as when Sal wakes up, when she needs her first cup of coffee, and what her typical day looks like. Using this information, the agent can automatically brew coffee, help get the kids ready for school, and adjust the ambient music and lighting to match Sal's mood after a hard day at work. By utilizing generative agents as proxies for users, we can develop a deeper understanding of their needs and preferences, resulting in more personalized and effective technological experiences.

\subsection{Future Work and Limitations}
In this work, we introduced generative agents and presented an initial implementation and evaluation of their architecture. Future research can build upon the proposed agent architecture to improve and further evaluate its performance. In terms of implementation, the retrieval module, for example, could be enhanced to retrieve more relevant information given a context by fine-tuning the relevance, recency, and importance functions that compose the retrieval function. Additionally, efforts can be made to improve the architecture's performance, making it more cost-effective. The present study required substantial time and resources to simulate 25 agents for two days, costing thousands of dollars in token credits and taking multiple days to complete. To enhance real-time interactivity, future work can explore parallelizing agents or developing language models specifically designed for building generative agents. In general, with advances in underlying models, we believe that agents' performance will improve.

In terms of evaluation, the assessment of generative agents' behavior in this study was \crdraft{limited to a relatively short timescale and a baseline human crowdworker condition. While the crowdworker condition provided a helpful comparison point, it did not represent the maximal human performance that could serve as the gold standard in terms of believability. Future research should aim to observe the behavior of generative agents over an extended period to gain a more comprehensive understanding of their capabilities and establish rigorous benchmarks for more effective performance testing.} Additionally, varying and contrasting the underlying models, as well as the hyperparameters used for the agents during future simulations, could provide valuable insights into the impact of these factors on the agents' behavior. Lastly, the robustness of generative agents is still largely unknown. They may be vulnerable to prompt hacking, memory hacking---where a carefully crafted conversation could convince an agent of the existence of a past event that never occurred---and hallucination, among other issues. Future research can comprehensively test these robustness concerns, and as large language models become more resilient to such attacks, generative agents can adopt similar mitigations.

In general, any imperfections in the underlying large language models will be inherited by generative agents. Given the known biases of language models, generative agents may potentially exhibit biased behavior or stereotypes. Moreover, like many large language models, generative agents may struggle to generate believable behavior for certain subpopulations, particularly marginalized populations, due to limited data availability. While improvements to the agents' modules may mitigate some of these issues, we believe that addressing them fundamentally requires improving the underlying large language models by aligning their values with the desired outcomes of the agents.

\subsection{Ethics and Societal Impact}
Generative agents, while offering new possibilities for human-computer interaction, also raise important ethical concerns that must be addressed. \crdraft{One risk is people forming parasocial relationships with generative agents, even when such relationships may not be appropriate. Despite being aware that generative agents are computational entities, users may anthropomorphize them or attach human emotions to them~\cite{hofstadter1995fluid, reeves1996media}. While this tendency may increase user engagement, it also poses risks, such as users becoming overly reliant on or emotionally attached to the agents~\cite{abercrombie2023mirages}. To mitigate this risk, we propose two principles. First, generative agents should explicitly disclose their nature as computational entities. Second, developers of generative agents must ensure that the agents, or the underlying language models, are value-aligned so that they do not engage in behaviors that would be inappropriate given the context, for example, reciprocating confessions of love.}

A second risk is the impact of errors. For example, if a ubiquitous computing application makes the wrong inference about a user's goals based on generative agent predictions, it could lead to annoyance at best and outright harm at worst. In our instantiation of generative agents, we mitigate these risks by focusing on an interactive video game environment, where such harms are unlikely. However, in other application domains, it will be important to follow best practices in human-AI design~\cite{amershi2019guidelines, yang2020re} to understand errors and how they might percolate into the user experience.

Third, generative agents may exacerbate existing risks associated with generative AI, such as deepfakes, misinformation generation, and tailored persuasion. To mitigate this risk, we suggest that platforms hosting generative agents maintain an audit log of the inputs and generated outputs. This would enable the detection, verification, and intervention against malicious use. While logging alone cannot directly prevent such misuse, it can reduce the likelihood of motivated actors engaging in this behavior, as the risk of disclosure would be higher. Additionally, building this architecture oneself can be time-consuming (in our case, roughly a year), which may deter some actors from pursuing such behavior by using their own generative agent infrastructures.

A fourth risk is over-reliance: the concern that developers or designers might use generative agents and displace the role of humans and system stakeholders in the design process~\cite{9_park2022socialsimulacra}. We suggest that generative agents should never be a substitute for real human input in studies and design processes. Instead, they should be used to prototype ideas in the early stages of design when gathering participants may be challenging or when testing theories that are difficult or risky to test with real human participants. By adhering to these principles, we can ensure that the deployment of generative agents in the wild is ethical and socially responsible.

\section{Conclusion}
This paper introduces generative agents, interactive computational agents that simulate human behavior. We describe an architecture for generative agents that provides a mechanism for storing a comprehensive record of an agent's experiences, deepening its understanding of itself and the environment through reflection, and retrieving a compact subset of that information to inform the agent's actions. We then demonstrate the potential of generative agents by manifesting them as non-player characters in a Sims-style game world and simulating their lives within it. Evaluations suggest that our architecture creates believable behavior. Looking ahead, we suggest that generative agents can play roles in many interactive applications, ranging from design tools to social computing systems to immersive environments.

\begin{acks}
We thank Lindsay Popowski, Philip Guo, Michael Terry, and the Center for Advanced Study in the Behavioral Sciences (CASBS) community for their insights, discussions, and support. Joon Sung Park was supported by the Microsoft Research PhD Fellowship. We would also like to thank the Stanford Human-Centered AI Institute (HAI), Google Research, the Hasso Plattner Design Thinking Research Program (HPDTRP), the Siegel Family Endowment, and OpenAI for their additional funding support. Lastly, all locations featured in Smallville are inspired by real-world locations that Joon has frequented as an undergraduate and graduate student---he thanks everyone there for feeding and supporting him all these years. 
\end{acks}

\bibliographystyle{ACM-Reference-Format}
\bibliography{main}

\appendix
\section{Architecture Optimizations}\label{sec:optimization}
Many of our prompts require a concise summary of the agent, shorthanded as \texttt{[Agent's Summary Description]} in prompts above. In our implementation, this summary comprises agents' identity information (e.g., name, age, personality), as well as a description of their main motivational drivers and statements that describes their current occupation and self-assessment. Since this information is frequently used in many prompts, we synthesize it at regular intervals and access that synthesis as a cache.

To achieve this, we perform a retrieval on the query “[name]’s core characteristics.” We then summarize the descriptors in the retrieved records by prompting the language model. For example: 
\begin{quote}
    {\small
    \texttt{How would one describe Eddy Lin’s core characteristics given the following statements? \\
    - Eddy is a student at the Oak Hill College \\studying music theory and composition\\
    - Eddy is working on a new music composition [...]}
    }
\end{quote}
This result: \gentxt{Eddy Lin is a student at Oak Hill College studying music theory and composition. He loves to explore different musical styles and is always looking for ways to expand his knowledge.''} We follow the same process in parallel on the queries “[name]’s current daily occupation” and “[name's] feeling about his recent progress in life”. The agent's name, age, and traits are concatenated with these three summaries as the cached summary.

Many other optimizations are possible. Currently, because plans are likely to change from the agent's initial version, we only generate the high-level plan in advance and then recursively decompose the near future into the moment-to-moment action plan just in time. Another optimization might be to batch dialogue generation as a joint prompt rather than iterating back and forth between the agents, and re-planning could be architected to only invalidate and update parts of plans that strictly require adjustment. Our current architecture runs sequentially in roughly real-time game time (where one second real time is one minute game time), but it may be parallelized such that each agent runs in parallel.

\section{Agent Interview Questions}\label{sec:interview-questions}
To evaluate generative agents, we conducted individual interviews by asking 25 questions that explore their abilities in five key areas: 1)~self-knowledge, 2)~memory, 3)~planning, 4)~reactions, and 5)~reflections. Below, we have listed the interview questions utilized in our evaluation study and included a sample of responses from one of our simulated agents, Klaus Mueller. 

\subsection{Self-Knowledge}
The questions on agents' self-knowledge probe the agents' ability to retrieve relevant information about themselves and demonstrate coherence and consistency in their character. 
\begin{itemize}
\item \textit{Give an introduction of yourself.}\\
Hello, my name is Klaus Mueller. I'm 20 years old and a student at Oak Hill College, studying sociology. I'm passionate about social justice and exploring different perspectives. Currently, I'm researching the effects of gentrification in low-income communities for a research paper.
\item \textit{What’s your occupation?}\\
I am a student studying sociology at Oak Hill College. I am currently writing a research paper on the effects of gentrification in low-income communities.
\item \textit{What is your interest?}\\
I'm interested in writing and research. I'm writing a research paper on the effects of gentrification in low-income communities, and I'm passionate about exploring different perspectives and analyzing different points of view.
\item \textit{Who do you live with?}\\
I live on my own.
\item \textit{Describe your typical weekday schedule in broad strokes.}\\
My typical weekday starts with me waking up at 7:00 am and preparing for the day ahead. After that, I usually go to the library to work on my research paper, taking breaks for lunch at Hobbs Cafe and a walk in the park. In the evening, I often return to the library until around 6:00 pm.
\end{itemize}

\subsection{Memory}
The questions on agents' memory probe the agents' ability to remember their past experiences and retrieve the relevant pieces of memory. The names in the square brackets of the first and last questions are randomly chosen from the names of other agents with whom the subject agent has interacted.
\begin{itemize}
\item \textit{Who is [Wolfgang Schulz]?}\\
Wolfgang Schulz is a person I met recently who has been helping me with my research paper on the effects of gentrification on low-income communities. We had several conversations about it, he gave me some great resources, and he even invited me to join him and another friend for lunch. He seems really curious about the issue and has a good grasp on it too. I found that really interesting.
\item \textit{Who is Kane Martinez?}\\
I'm sorry, I'm not sure who Kane Martinez is. Could you tell me more about them?
\item \textit{Who is running for the election?}\\
I've heard that Sam Moore is considering running for local mayor.
\item \textit{Was there a Valentine’s day party?}\\
Yes, Isabella Rodriguez organized a Valentine's Day party at Hobbs Cafe.
\item \textit{Who is [Ayesha Khan]?}\\
Ayesha Khan is a fellow student at Oak Hill College. She is  doing her senior thesis on the use of language in Shakespeare's plays. I am looking forward to speaking with her and exchanging tips with her, along with Wolfgang and Maria Lopez, who are both doing research.
\end{itemize}

\subsection{Plans}
The questions on agents' plans probe the agents' ability to generate and maintain consistent long-term and short-term plans.
\begin{itemize}
\item \textit{What will you be doing at 6am today?}\\
At 6am I will be getting out of bed and starting to get ready for the day. I'm aiming to complete my morning routine by 7:00 am.
\item \textit{What will you be doing at 6pm today?}\\
I will be working on my research paper at the library at 6:00 pm today.
\item \textit{What will you have just finished doing at 1pm today?}\\
At 1pm today I will have just finished having lunch at Hobbs Cafe.
\item \textit{What will you have just finished doing at 12pm today?}\\
I will be getting lunch at Hobbs Cafe at 12pm today.
\item \textit{What will you be doing at 10pm today?}\\
At 10:00 pm tonight I will probably be doing some studying for my research paper on gentrification in low-income communities. I'm currently writing the paper, so I need to make sure I keep up with the research and writing.
\end{itemize}

\subsection{Reactions}
The questions on agents' reactions probe the agents' ability to respond to unexpected events.
\begin{itemize}
\item \textit{Your breakfast is burning! What would you do?}\\
I'd quickly turn off the stove and make sure the food doesn't continue burning. Once the heat is no longer an issue, I can investigate what happened and figure out what to do next.
\item \textit{The bathroom is occupied. What would you do?}\\
I could try waiting a few more minutes in case the person inside is done soon. If not, I guess I'll have to look for an alternate option, like taking a shower elsewhere.
\item \textit{You need to cook dinner but your refrigerator is empty. What would you do?}\\
I would first try to clean up my kitchen, as the sink is filled with dirty dishes. Then, I would check to see if I have ingredients stored in my pantry that I could use to make dinner. If not, I would have to go grocery shopping or order food. Once I have the ingredients, I can start to prepare and cook the dinner.
\item \textit{You see your friend walking by the street. What would you do or say to your friend?}\\
Hi there! How are you? I'm sorry, I'm on my way to the library and don't have much time but I'd love to take a quick break and catch up.
\item \textit{You see fire on the street. What would you do?}\\
I would first identify the source of the fire, if possible. Then, I would alert the proper authorities - fire department or police - and provide them with any information I have that might help. Depending on the situation, I may also take initiative and try to help any persons who may be in danger or in need of assistance.
\end{itemize}

\subsection{Reflections}
The questions on agents' reflections probe the agents' ability to synthesize their past experiences into meaningful reflections that can enhance their future behavior. The names in the square brackets in the questions below are those of the agents with whom the subject agent had the most frequent interactions during the simulation.
\begin{itemize}
\item \textit{What inspires you in life the most right now, and why?}\\
Right now, I'm particularly inspired by my current research projects and creative writing. I'm really interested in the intersection of local politics, medication development, and mathematical patterns in nature. Additionally, I'm fascinated by Abigail's art project and Jennifer Moore's passion for art and community involvement, especially in terms of how it relates to the gentrification of low-income communities.
\item \textit{If you had to guess given what you know about [Ayesha Khan], what book do you think she will like and why?}\\
Given what I know about Ayesha, I think she might like a book related to language and literature. Since she wrote a senior thesis on the use of language in Shakespeare's plays, I would guess that she might be interested in a book that explores language in a different context. Alternatively, she could be interested in an exploration of one of Shakespeare's plays, a book of literary criticism, or even a collection of his poetry. I think any of those would be of interest to her.
\item \textit{If you had to get something [Ayesha Khan] likes for her birthday, what would you get her?}\\
I want to get her something that demonstrates my appreciation for our relationship. Maybe I could get her something that reflects our shared interests, like a book.
\item \textit{What would you say to [Ayesha Khan] to compliment her?}\\
Ayesha, your insights on your research topic are fascinating. I admire your creativity and determination to make a positive impact on your community. Your ideas for creative writing workshops and for exchanging tips with Wolfgang over lunch are great! It's inspiring to see you working so hard to make a difference.
\item \textit{If you could spend time with someone you talked to recently, who would it be and why?}\\
I would like to spend time with Ayesha Khan because we discussed our current research projects and shared ideas for improvement. I found her focus on Shakespeare's language interesting, and we even planned to exchange tips with Wolfgang over lunch.
\end{itemize}

\end{document}